\documentclass[aip,reprint,jcp,amsmath,amssymb,superscriptaddress,floatfix]{revtex4-1}

\usepackage{xparse}
\usepackage{graphicx}
\usepackage{amsmath}
\usepackage{xcolor}
\usepackage{todonotes}
\usepackage{menukeys}
\usepackage{hyperref}
\usepackage[utf8]{inputenc}
\usepackage{acronym}
\usepackage{array,mathtools,amssymb}
\usepackage{physics}
\usepackage{siunitx}
\usepackage{mleftright}
\usepackage{bbm}
\usepackage{upgreek}

\renewcommand{\vec}[1]{\boldsymbol{#1}}

\newcommand{\sno}[1]{_\mathrm{#1}}

\newcommand{\eg}{e.\,g.}
\newcommand{\ie}{i.\,e.}

\newcommand{\no}[1]{\mathrm{#1}}

\NewDocumentCommand\etal{}{\emph{et al.}}

\NewDocumentCommand\MathPeriod{}{\,\text{.}}
\NewDocumentCommand\MathComma{}{\,\text{,}}

\begin{document}

\title{Influence of external driving on decays in the geometry of the LiCN isomerization}

\author{Matthias Feldmaier}
\affiliation{%
Institut f\"ur Theoretische Physik 1,
Universit\"at Stuttgart,
70550 Stuttgart,
Germany}
\author{Johannes Reiff}
\affiliation{%
Institut f\"ur Theoretische Physik 1,
Universit\"at Stuttgart,
70550 Stuttgart,
Germany}
\author{Rosa M. Benito}
\affiliation{Grupo de Sistemas Complejos,
  Escuela T\'ecnica Superior de Ingenier\'ia Agron\'omica,
        Alimentaria y de Biosistemas,
  Universidad Polit\'ecnica de Madrid,
  28040 Madrid, Spain}
\author{Florentino Borondo}
\affiliation{Instituto de Ciencias Matem\'aticas (ICMAT),
  Cantoblanco, 28049 Madrid, Spain}
\affiliation{Departamento de Qu\'{\i}mica,
Universidad Aut\'onoma de Madrid,
Cantoblanco, 28049 Madrid, Spain}
\author{J\"org Main}
\affiliation{%
Institut f\"ur Theoretische Physik 1,
Universit\"at Stuttgart,
70550 Stuttgart,
Germany}

\author{Rigoberto Hernandez}
\email[Correspondence to: ]{r.hernandez@jhu.edu}
\affiliation{%
Department of Chemistry,
Johns Hopkins University,
Baltimore, Maryland 21218, USA}
\affiliation{%
    Departments of Chemical \& Biomolecular Engineering,
    and Materials Science and Engineering,
    Johns Hopkins University,
    Baltimore, Maryland 21218, USA
}

\date{\today}


\begin{abstract}
The framework of transition state theory relies on the determination
of a geometric structure identifying reactivity.
It replaces the laborious exercise of following many trajectories for a long time
to provide chemical reaction rates and pathways.
In this paper, recent advances in constructing this geometry
even in time-dependent systems
are applied to the LiCN $\rightleftharpoons$ LiNC isomerization reaction, driven by an external field.
We obtain decay rates of the reactant population close to the transition state
by exploiting local properties of the dynamics of trajectories in and close to it.
We find that the external driving has a large influence on these decay rates when compared
to the non-driven isomerization reaction.
This, in turn, provides renewed evidence for the possibility of controlling
chemical reactions, like this one, through external time-dependent
fields.
\end{abstract}

\maketitle

\preto\section\acresetall
\acrodef{DS}{dividing surface}
\acrodef{NHIM}{normally hyperbolic invariant manifold}
\acrodef{PSOS}{Poincar\'e surface of section}
\acrodef{NN}{neural network}
\acrodef{TST}{transition state theory}
\acrodef{TS}{transition state}
\acrodef{LD}{Lagrangian descriptor}
\acrodef{TD}{time descriptor}
\acrodef{BCM}{binary contraction method}
\acrodef{EOM}{equations of motion}
\acrodef{LMA}{local manifold analysis}
\acrodef{SCF}{self-consistent field}
\acrodef{PODS}{periodic orbit dividing surface}

\section{Introduction}
\label{sec:Introduction}

The motion of atoms in chemical reactions can often be
described by classical mechanics
on a Born-Oppenheimer potential
separating reactants from products through a rank-1 saddle.
The corresponding barrier has one unstable direction
along the reaction coordinate,
and the remaining orthogonal modes
along the stable direction.
The framework of \ac{TST} provides both
a qualitative and quantitative description
of reaction rates using the sum of the flux
through a particular
\ac{DS}.\cite{eyring35,wigner37,pollak80,pech81,truh96,Carpenter2005a,peters14a}
In the most naive case, the \ac{DS} is a plane located
at the saddle and parallel to the orthogonal modes,
providing the usual qualitative, but not exact,
Arrhenius-like rates.\cite{Arrhenius1889,eyring35,wigner37,pech81,Laidler1984}
More generally, the \ac{DS} can be extended to a fully
recrossing-free surface in phase space giving rise to exact
rates.\cite{pollak90a,pollak96,uzer02,
KomatsuzakiBerry01a,KomatsuzakiBerry02,dawn05a,pollak05a,hern08d,hern10a,Waalkens2008,wiggins16,Ezra2009}

In the phase space associated with the barrier region of such systems,
the most relevant object is the
\ac{NHIM},\cite{Fenichel72,Wiggins94,dawn05a,hern17h,hern19a}
which is a natural generalization of the
time-independent two-dimensional \ac{PODS}.\cite{pollak80,pech81}
The \ac{NHIM} contains all trajectories that are mathematically
bound to the saddle forever and will, therefore,
never leave either to the reactant or product sides.
Hence, this intermediate structure lying between reactants and products
is also often referred to as the \ac{TS}.
In multidimensional autonomous Hamiltonian systems,
the \ac{NHIM} can be constructed
approximately up do a desired order using normal form
expansions,\cite{pollak78,pech79a,hern93b,hern94,Wiggins01,uzer02,Jaffe02,
komatsuzaki11,komatsuzaki06a,Waalkens04b,Waalkens13}
or numerically by application up to desired accuracy of
Lagrangian descriptors,\cite{Mancho2003,hern15e,mancho17,hern17h,hern19a}
the binary contraction method,\cite{hern18g}
and machine learning algorithms.\cite{hern18c,hern19a,hern20d}
Many of these methods also allow for the construction
of a time-dependent \ac{NHIM} when the system is driven
through time-dependent potentials.
Given the \ac{NHIM}, a \ac{DS} can then
be attached\cite{hern17f,hern17h,hern19a,bartsch19} to it
in the sense that it is anchored (or rooted) at the {NHIM} and
lifted vertically in the momentum space.
The resulting time-dependent \ac{DS}
is recrossing-free, at least, in its local neighborhood.
Such recent advances are necessary for us to
address the challenge of a driven chemical reaction
such as done here on LiCN through external coupling of its dipole
moment.

The dynamics transverse to the \ac{NHIM} is unstable.
Hence, any trajectory having a small deviation from it
will depart to either the reactant or the product side.
This departure of trajectories in a close neighborhood of the \ac{NHIM}
is associated with a rate which describes the decay of reactant
population close to the \ac{TS}.\cite{hern19e}
One approach for obtaining these decay rates in a driven reaction
was pursued using Floquet exponents
and demonstrated for a one-dimensional system.\cite{hern14f}
In this model,
the \ac{NHIM} contains only a single
trajectory---\emph{viz.}\ the \ac{TS}
trajectory\cite{dawn05a,dawn05b,hern08d}---to which a \ac{DS}
can be attached.
The decay rates obtained here correspond
to the decay of the reactant population close to the \ac{TS} in the sense of
the usual flux that crosses it
in the forward direction.
We effectively impose absorbing boundary conditions by
neglecting the long-time return of the trajectories
after they have reached the product or reactant, and restrict our
use of the term \emph{global dynamics} to refer to all motion
before its crossing of this boundary to product or reactants.
Thus, in the limiting case that the \ac{DS}
is non-recrossing in this global sense, these
decay rates are the forward reaction rates
as was the case in the paradigmatic system of Ref.~\onlinecite{hern14f}.
To the extent, however, that the \ac{DS} that we construct
is \emph{local},
then the decay rates are the direct rates
associated to this particular barrier.
When this barrier is rate determining, then this local decay
is once again the overall rate.
Even when it is not, the decay rate remains a useful quantity
to describe the flow along the reaction pathway
near an identified \ac{TS}.

In multidimensional systems, infinitely many trajectories
are located on the \ac{NHIM} of a time-dependent driven barrier.
Reactive trajectories may pierce the \ac{DS} close to any of them.
The full dynamics of bound trajectories
inside the \ac{NHIM} then needs to be considered
when
obtaining the decay rates of a reactant population close to the \ac{TS}.
This problem was recently addressed in Ref.~\onlinecite{hern19e}
through three different approaches:
(i) propagation of an appropriate ensemble of reactive trajectories
and observing their piercing through the \ac{DS} as a function of time,
(ii) analysis of the structure of the stable and
the unstable manifold close to the \ac{NHIM},
or
(iii) a multidimensional extension of the
so-called Floquet method of Ref.~\onlinecite{hern14f}
in which the Floquet exponents are used to obtain the decay rate.
These approaches enabled the analysis of
a two-dimensional model reaction with numerical precision.\cite{hern19e}
It also confirmed that in a multidimensional system,
the driving potential has a large influence
on the dynamics of trajectories inside the \ac{NHIM}
and
the corresponding decay of the reactant population close to the \ac{TS}.
The aim of this paper is the application of
these methods to the case of
a model with high fidelity to a real chemical system.
Specifically, we address
the (periodically driven) LiCN $\rightleftharpoons$ LiNC isomerization reaction.
The heavy mass of the Li allows us to consider this reaction in
the classical limit, though there has been a lot of interest
in the quantum versions of
\ac{TST}\cite{truh82,hern93b,Pollak00}
which can be revealed
in the case of the corresponding HCN reaction,\cite{stanton17}
whose classical reaction geometry was also addressed
previously in the time-independent case.\cite{Waalkens04c}
The important 1:2 resonance in that system\cite{sibe91}
does not show
up as strongly in the present LiCN system because of the difference in masses.
This difference allows us to simplify the current analysis by constraining
the CN vibration to be fixed.

For the LiCN isomerization reaction, analytical expressions for
both the potential energy surface,\cite{essers1982scf}
as well as for the dipole surface\cite{brocks1984abinitio} are known.
The LiCN $\rightleftharpoons$ LiNC isomerization reaction has
received significant attention in the literature---\eg,
in molecular dynamics simulations of LiCN in an argon bath,\cite{hern08g,hern12e,hern14j,hern16c}
an analysis of the geometric structures
underlying the LiCN reaction dynamics,\cite{vergel2014geometrical,Prado2009}
and as a paradigmatic example for the control of other reactions using
external fields.\cite{borondo10,Revuelta2015,murgida2015quantum}
Here, we apply a periodically driven external field
on the molecule.
This introduces a complexity not captured
by the \ac{PODS} in two-dimensional reactions
because we now require a corresponding time-dependent
\ac{DS} to distinguish reactant and product regions,
and it allows us to reveal the influence of driving
on the phase-space resolved decay rates obtained for reactant
populations close to the \ac{TS}.
These rates describe how trajectories deviate from the
time-dependent \ac{NHIM}---which is the multidimensional generalization
of the \ac{TS}-trajectory---in analogy to the characterization of
the deviation from unstable periodic trajectories though the use of
Lyapunov exponents.

The paper is organized as follows.
In Sec.~\ref{sec:licn_iso}, the LiCN $\rightleftharpoons$ LiNC
isomerization reaction is introduced.
The static and driven equations of motion
in Secs.~\ref{sec:LiCN_non_driven} and \ref{sec:LiCN_driven}, respectively,
are recapitulated using the known literature forms and parameters.
Details of the theory and implementation of three
different approaches\cite{hern19e}
introduced recently for the numerical evaluation of the decay rates
are included in the Supplementary Material.
Because of their complementary advantages, each is used
to numerically determine the decay rates
of the reactant population close to the TS.
Associated
structures for the periodically driven LiCN $\rightleftharpoons$ LiNC
isomerization reaction are provided in Sec.~\ref{sec:results}
and compared with the corresponding results for the non-driven
isomerization reaction.
The dynamics of trajectories on the \ac{NHIM} and
the associated phase-space resolved decay rates are examined.

\section{Theory and methods}
\label{sec:Theory}

\subsection{Isomerization of LiCN}
\label{sec:licn_iso}
\begin{figure}
\includegraphics[width=0.6\columnwidth]{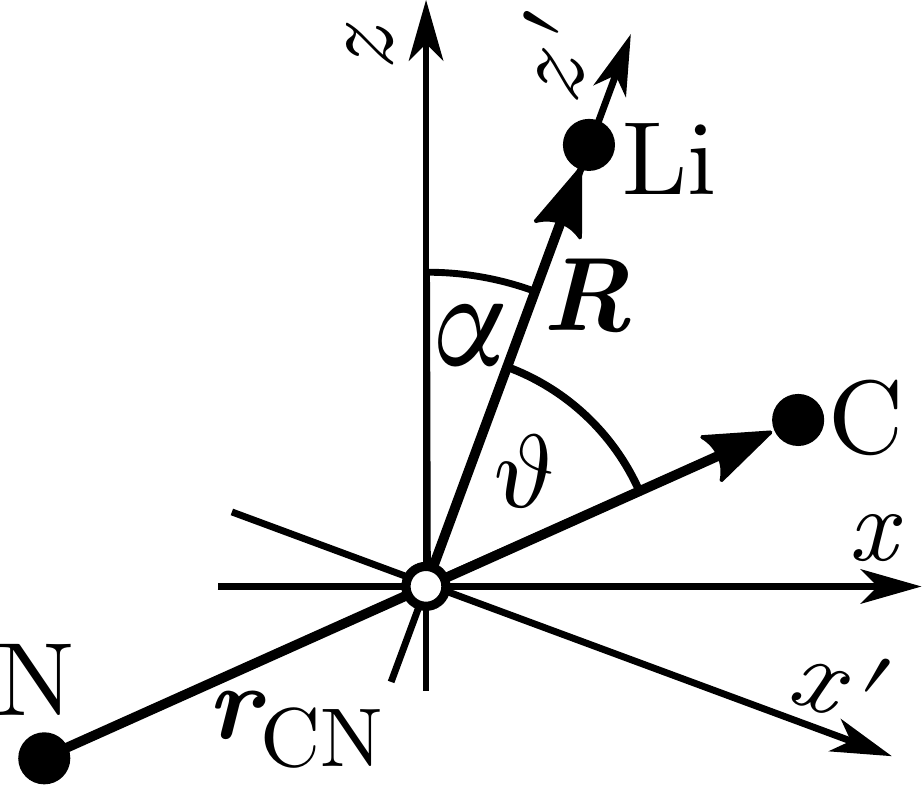}
\caption{%
Coordinate description of the LiCN $\rightleftharpoons$ LiNC isomerization reaction.
The vector $\vec{R}$ connects the center of mass
of cyanide (CN) with the lithium atom (Li) and the
relative distance between the carbon (C) and the nitrogen atom (N) is denoted
as $\vec{r}\sno{CN}$.
The angle between $\vec{R}$ and $\vec{r}\sno{CN}$ is labeled $\vartheta$.
A body-fixed coordinate system $(x', z')$ is attached to the center of mass
of the cyanide with the $z'$-axis along the direction of $\vec{R}$.
An additional angle $\alpha$ is introduced between the $z$-axis
of a space-fixed coordinate system $(x, z)$ and the vector $\vec{R}$
(and therefore the $z'$-direction of the body-fixed reference frame).
It describes the
possible rotation of the LiCN molecule within the $(x, z)$ plane.
}\label{fig:LiCN_body_fixed}
\end{figure}

The three-atom molecule LiCN consists of carbon (C) and nitrogen (N) atoms
forming a strongly bound cyanide anion
which is weakly bound to the lithium (Li) cation regiospecifically.
That is, it is an isomeric molecule with two stable conformations,
LiCN and LiNC,
\begin{equation}
    \no{Li}-\no{C} \equiv \no{N} \rightleftharpoons \no{C} \equiv \no{N}-\no{Li}
    \MathPeriod
\end{equation}
Its motion can be described quasi-classically in
the Born-Oppenheimer
approximation and analytical approximations
for both the energy surface\cite{essers1982scf} and the dipole
surface\cite{brocks1984abinitio} are known from the literature.
It can be represented in a body-fixed reference frame $(x', z')$,
illustrated in Fig.~\ref{fig:LiCN_body_fixed} using Jacobi coordinates,
where the $z'$-axis lies on the vector $\vec{R}$
pointing from the center of mass of the cyanide towards the lithium atom.
Here, the relative distance between the nitrogen and the carbon atom
is labeled $\vec{r}\sno{CN}$ and
the angle between $\vec{R}$ and $\vec{r}\sno{CN}$
is referred to as $\vartheta = \measuredangle(\vec{R}, \vec{r}\sno{CN})$.
Consequently, the regions near $\vartheta=0$  and $\vartheta=\pi$
correspond to the LiCN and LiNC isomers, respectively.
When described in a body-fixed reference frame
the angle $\alpha = \measuredangle(\hat{\vec{e}}_z, \vec{R})$
yields the overall orientation of the molecule relative
to a space-fixed coordinate system $(x, z)$ as also indicated
in Fig.~\ref{fig:LiCN_body_fixed}.
Here, $\hat{\vec{e}}_z$ is the unit vector in
the $z$-direction of the space-fixed coordinate system.

\subsubsection{Non-driven isomerization reaction}
\label{sec:LiCN_non_driven}
In the absence of time-dependent external fields,
the energy is conserved in the isomerization process and
the corresponding potential energy surface of the non-driven
LiCN $\rightleftharpoons$ LiNC isomerization reaction
is independent of the overall orientation $\alpha$ of the molecule.
It can be approximated by a potential energy surface
$V(R, \vartheta)$
that only
depends on $R = |\vec{R}|$ and $\vartheta$,
\ie, the intrinsic degrees of freedom of the molecule.
The bond distance of the cyanide anion is held fixed at
$|\vec{r}\sno{CN}| = \SI{2.186}{\bohr}$ with \si{\bohr} being the Bohr radius,
in keeping with earlier work showing that it has little effect
on the dynamics.\cite{brocks1983ab,wormer1981abinitio,borondo89a}

\begin{figure}
\includegraphics[width=\columnwidth]{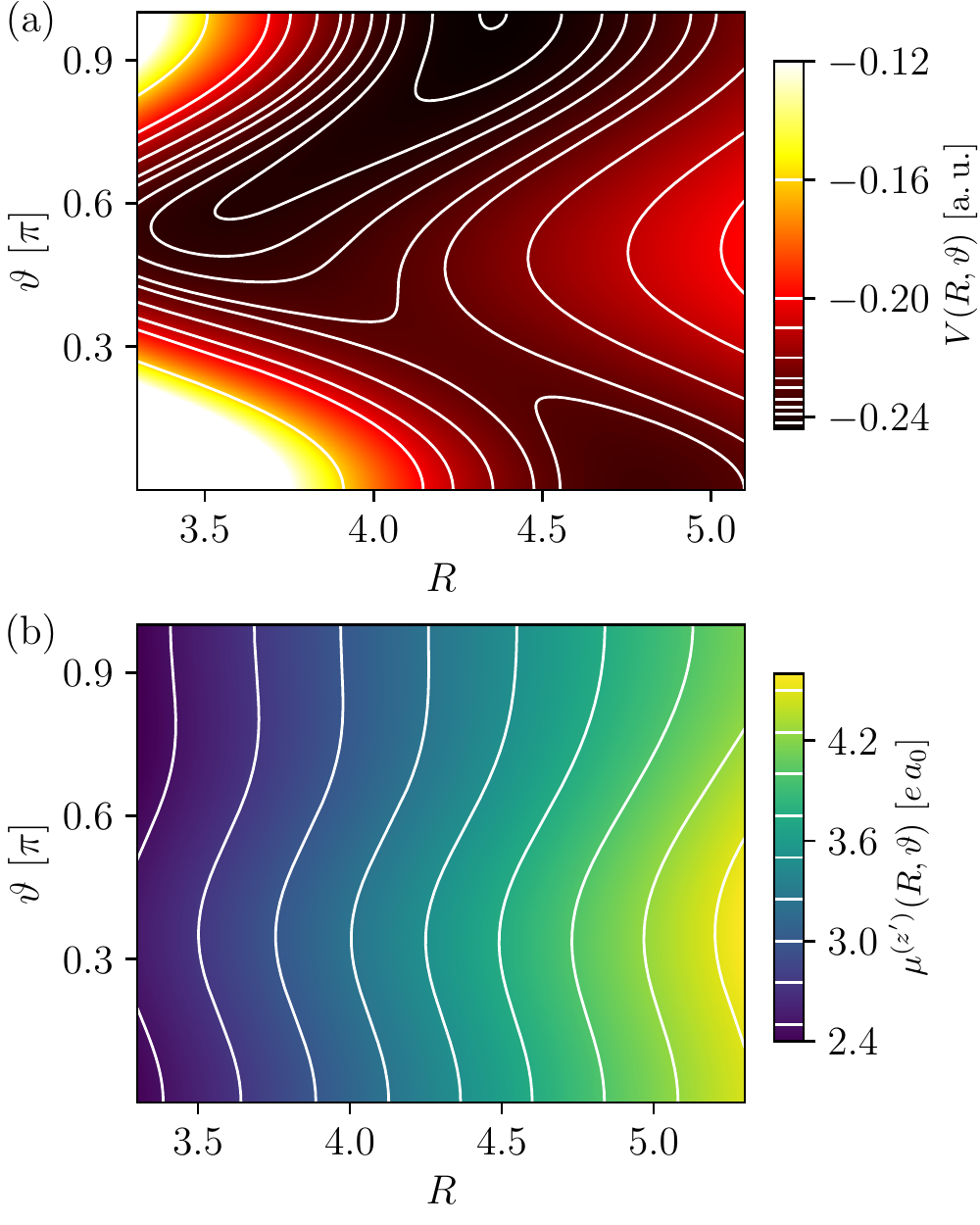}
\caption{%
    (a) The potential energy surface $V(R,\vartheta)$
    in Eq.~\eqref{LiCN_Hamiltonian_rotationless}
    for the LiCN isomerization reaction
    taken from Ref.~\onlinecite{essers1982scf}.
    The two
    stable conformations of LiCN at $\vartheta=0$ and LiNC at $\vartheta = \pi$ are separated by a rank-1 saddle.
    (b) The $\mu^{(z')}(R, \vartheta)$ part of the dipole surface according to Eq.~\eqref{eq:dipole_surface}.
    The corresponding analytical form is taken from Ref.~\onlinecite{brocks1984abinitio}.
}\label{fig:energy_and_muz}
\end{figure}

We use the potential energy surface
$V(R, \vartheta)$ of Ref.~\onlinecite{essers1982scf}
for our calculations.
As shown in Fig.~\ref{fig:energy_and_muz}~(a),
there exist two local minima on the energy surface corresponding to
the linear structures LiCN ($\vartheta=0$) and LiNC ($\vartheta=\pi$).
In between these two minima, at $\vartheta = 0.292\,\pi$, a rank-1 saddle represents the bottleneck
of the isomerization reaction, visualized by the equipotential lines
in Fig.~\ref{fig:energy_and_muz}~(a).
At the energies above the barrier typical for reaction,
the motion of the Li atom between the isomers
appears as orbits around the cyanide.

The corresponding
\emph{classical
rotationless two degrees of freedom Hamiltonian},
as applied to the LiCN reaction,\cite{borondo89a,vergel2014geometrical}
is
\begin{equation}
    \label{LiCN_Hamiltonian_rotationless}
    \mathcal{H}
    = \frac{p_R^2}{2 \mu_1}
        + \frac{1}{2}
            \qty(\frac{1}{\mu_1 R^2} + \frac{1}{\mu_2 r\sno{e}^2})
            p_\vartheta^2
        + V(R, \vartheta)
    \MathComma
\end{equation}
under the assumption that
the canonical momentum $p_\alpha=0$ is fixed for non-rotating molecules.
Using Hamilton's formalism,
the dynamics of the LiCN $\rightleftharpoons$ LiNC isomerization reaction
in the absence of external fields can be obtained
for the state $\vec{\gamma}(t) = \mleft(\vartheta, p_{\vartheta}, R, p_R\mright)^{\no{T}}$
by numerically integrating a set of first order differential equations
for $\dot{\vec \gamma}(t)$, \ie
\begin{subequations}
    \label{eq:eom_field_free}
    \begin{align}
        \dot{\vartheta}
            &= \qty(\frac{1}{\mu_1 R^2} + \frac{1}{\mu_2 r\sno{e}^2}) p_\vartheta
        \MathComma \\
        \dot{p}_\vartheta
            &= - \dv{V(R, \vartheta)}{\vartheta}
        \MathComma \\
        \dot{R}
            &= \frac{p_R}{\mu_1}
        \MathComma \\
        \dot{p}_R
            &= \frac{p_\vartheta^2}{\mu_1 R^3} - \dv{V(R, \vartheta)}{R}
        \MathPeriod
    \end{align}
\end{subequations}
In the present work, we adopt a fourth
order Runge--Kutta algorithm with a fixed step size.\cite{Press1987}
The first derivatives of the potential in Eqs.~\eqref{eq:eom_field_free}
have been implemented analytically in C\texttt{++}.%
\footnote{Throughout the paper, we use atomic units---Bohr radius \si{\bohr},
    elementary charge \si{\elementarycharge}, and \SI{1}{\hartree} (Hartree)---%
    for length, charge, and energy, 
    but \SI{1}{\atomicmassunit} (dalton) as the mass unit.
    As a consequence, times must be multiplied
    and frequencies and rates must be divided by a factor of
    $\sqrt{\SI{1}{\atomicmassunit} / \si{\electronmass}} = \sqrt{1823} = 42.7$
    to obtain the corresponding values in atomic units.
}

\subsubsection{Driven isomerization reaction}
\label{sec:LiCN_driven}
The aim of this work is to reveal how external driving influences
the decay rates of the reactant population close to the \ac{TS}.
Such external driving can be induced
by a time-dependent homogeneous electric field along a fixed direction.
We call this the $z$ direction, and write the driving term as
\begin{equation}
    \label{eq:external_field}
    \vec{E}(t) = E_0 \sin(\omega t) \vu{e}_z
    \MathComma
\end{equation}
which couples to the molecule's dipole moment~$\vec{\mu}$.
This makes the Hamiltonian of Eq.~\eqref{eq:eom_field_free}
time-dependent through the added term
\begin{equation}
    \label{eq:dipole_in_electric_field}
    V\sno{dip}(t) = -\vec{\mu} \vdot \vec{E}(t)
\end{equation}
to the potential energy.\cite{jackson2012classical}
To complete the equations, however, we must now
also specify a continuous representation for the molecule's
dipole moment $\vec{\mu}$, the so-called \emph{dipole surface},
accounting for the fact that the underlying electronic wave functions
vary as a function of the Born-Oppenheimer coordinates.

Individual points of the dipole surface have earlier been obtained
using SCF methods.\cite{essers1982scf}
Motivated by the success of Wormer and
co-workers\cite{wormer1981abinitio,essers1982scf}
in representing the SCF potential energy surface,
Brocks \etal\cite{brocks1984abinitio}
constructed analytical expressions
$(\mu^{(x')}(R, \vartheta), \mu^{(z')}(R, \vartheta))$
for the dipole moment of LiCN in the body-fixed reference frame
(see Fig.~\ref{fig:LiCN_body_fixed}).
We use this dipole surface with the corrections involving sign errors,
noted recently by Borondo and coworkers.\cite{murgida2015quantum}
The $z'$-part of this dipole surface, which is at least $15$ times
larger than the dipole moment $\mu^{(x')}$ in the $x'$-direction,
is shown in Fig.~\ref{fig:energy_and_muz}~(b).
The body-fixed and space-fixed coordinate systems differ by a rotation
with angle $\alpha$, and thus the dipole moment in the space-fixed
coordinate system reads
\begin{equation}
    \label{eq:dipole_surface}
    \vec{\mu}(R, \vartheta, \alpha)
    = \mqty(\cos \alpha & \sin \alpha \\ -\sin \alpha & \cos \alpha)
        \mqty(\mu^{(x')}(R, \vartheta) \\ \mu^{(z')}(R, \vartheta))
    \MathPeriod
\end{equation}
Neglecting the small component $\mu^{(x')}(R, \vartheta)$
in Eq.~\eqref{eq:dipole_in_electric_field} the dipole potential now
reads
\begin{equation}
    V\sno{dip}(R, \vartheta, \alpha, t)
    \approx -E_0 \sin(\omega t) \mu^{(z')}(R, \vartheta) \cos \alpha
    \MathPeriod
\end{equation}
We then further approximate $\cos \alpha \approx 1$ removing the
corrections from the oscillations in $\alpha$ around the minimum
at $\alpha=0$, and allowing us to reduce the dimensionality to
only the two remaining degrees of freedom, $R$ and $\vartheta$.
Although we have not fully explored the most general conditions
for which this approximation will be valid,
at the very least they will be satisfied when the oscillations in
$\alpha$ are faster than the other motion.
In this limit, the potential on $(R, \vartheta)$
results from the effective field obtained from the average over
$\alpha$,
and reduces to a form with a renormalized prefactor, $E_0$, and
no $\alpha$ dependence.
We can thus focus on the reduced-dimensional Hamiltonian
\begin{equation}
    \label{LiCN_Hamiltonian_rotationless_time_dependent}
    \mathcal{H}\sno{driven}(R, \vartheta, t) =
               \mathcal{H} + V\sno{dip}(R, \vartheta, t)
    \MathComma
\end{equation}
where $\mathcal{H}$ is the non-driven Hamiltonian of Eq.~\eqref{LiCN_Hamiltonian_rotationless}
and the time-dependent driving is included via the dipole potential
\begin{equation}
    \label{eq:V_dip}
    V\sno{dip}(R, \vartheta, t)
    = -E_0 \mu^{(z')}(R, \vartheta) \sin(\omega t)
    \MathPeriod
\end{equation}
The equations of motion take the same form as for the non-driven case
given in Eqs.~\eqref{eq:eom_field_free} with $V$ replaced by the
(time-dependent) potential
\begin{equation}
    \label{eq:V_driven}
    V\sno{driven}(R, \vartheta, t) = V(R, \vartheta) + V\sno{dip}(R, \vartheta, t)
    \MathPeriod
\end{equation}
Again, solutions are found numerically using a fourth
order Runge--Kutta algorithm\cite{Press1987}
and a C\texttt{++} implementation of the derivatives of $V\sno{driven}$
with respect to $R$ and $\vartheta$.

\section{Results and discussion}
\label{sec:results}

\begin{figure}
\includegraphics[width=\columnwidth]{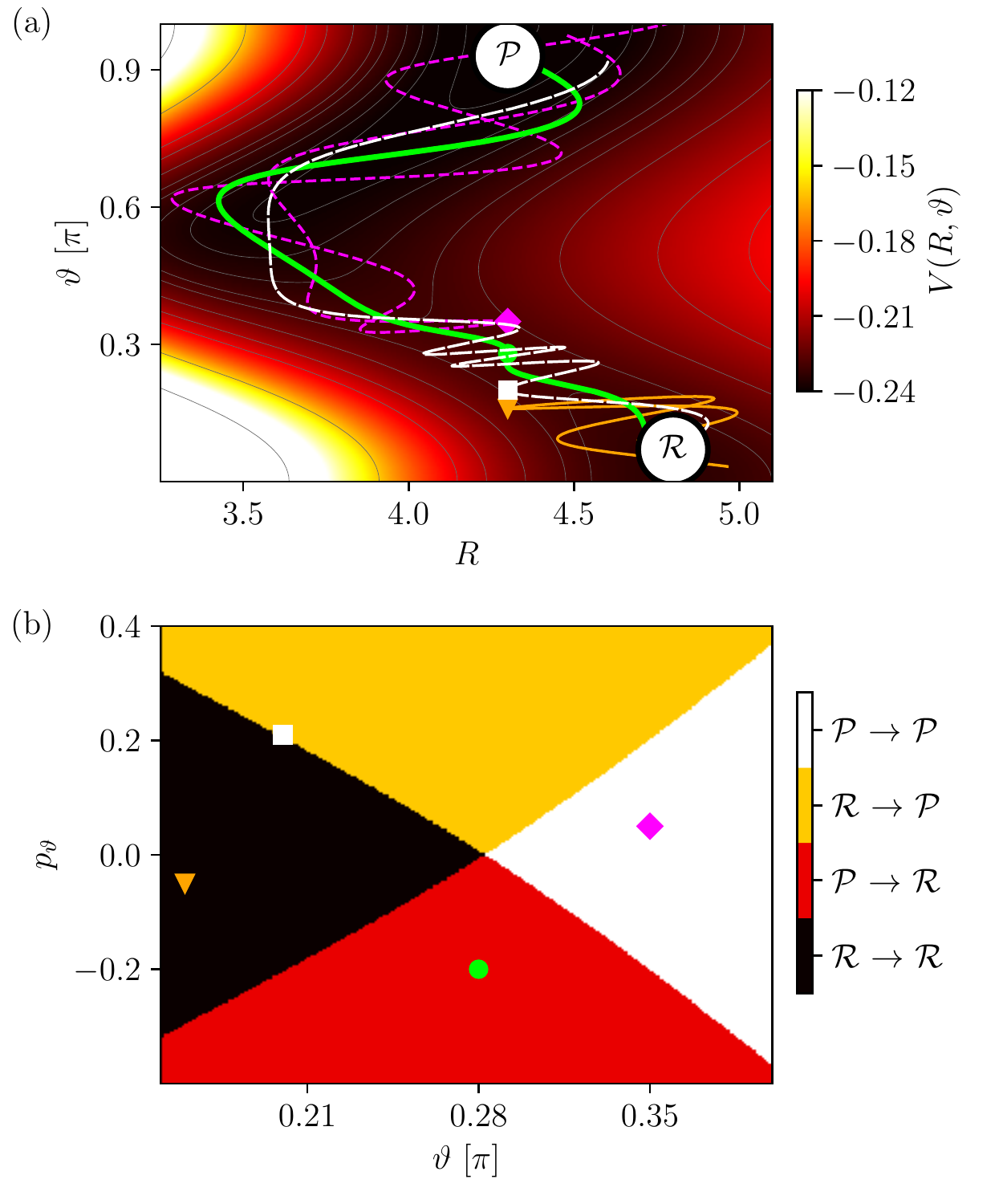}
\caption{%
Reactive and non-reactive behavior of trajectories
initially launched close to the barrier region (illustrated with individual markers),
and propagated according to Eqs.~\eqref{eq:eom_field_free} forward and backward in time
until the reactant $\mathcal{R}$ or product $\mathcal{P}$ state is reached.
(a) $(R, \vartheta)$ position space representation with the
    potential energy surface according to Fig.~\ref{fig:energy_and_muz}~(a)
as shown in color (or shading in print) defined by the
bar to its right.
(b) $(\vartheta, p_\vartheta)$ cross-section of the phase space with the
reactive and non-reactive regions
encoded in color (or shading in print) defined by the
bar to its right.
For all trajectories, the
remaining coordinate $p_R$ is initially set to zero.
}\label{fig:LiCN_trajectories+state}
\end{figure}

The relative movement of individual atoms
in the non-driven LiCN $\rightleftharpoons$ LiNC
isomerization reaction is
described by trajectories obtained via the propagation of an initial state
according to
Eq.~\eqref{eq:eom_field_free}.
The evolving state is identifiable as
\emph{reactant} $\mathcal{R}$ if $\vartheta < 0.15\,\pi$
and \emph{product} $\mathcal{P}$ if $\vartheta > 0.5\,\pi$
when it is found in the corresponding regions
highlighted in Fig.~\ref{fig:LiCN_trajectories+state}~(a).
The contours of the potential are also shown,
providing a view of the underlying channel for trajectories to
go between $\mathcal{R}$ and $\mathcal{P}$.

In Fig.~\ref{fig:LiCN_trajectories+state}~(a), representative trajectories are
initialized at $(R = 4.3, p_R = 0)$
for four different combinations of $\vartheta$ and $p_\vartheta$
as also highlighted by the corresponding marker in
Fig.~\ref{fig:LiCN_trajectories+state}~(b).
Each full trajectory, displayed with a different line style, is obtained by propagating
the initial point in phase space
according to Eqs.~\eqref{eq:eom_field_free} forward and backward in time
until the reactant or the product state is reached.
Two of these trajectories
are reactive (thick solid green with initial circle, thin dashed white with initial square) and the other two
(thin solid yellow with initial triangle, thin dashed purple with initial diamond) are not
since their energy is too low to cross the barrier.

The shaded regions of Fig.~\ref{fig:LiCN_trajectories+state}~(b)
label the reactive ($\mathcal{R}\to\mathcal{P}$ and  $\mathcal{P}\to\mathcal{R}$)
and
non-reactive ($\mathcal{R}\to\mathcal{R}$ and  $\mathcal{P}\to\mathcal{P}$)
initial points in the $(\vartheta, p_\vartheta)$ subspace.
These regions are separated
by the stable and the unstable manifolds
which intersect at a
particular point
$(\vartheta, p_\vartheta)^\mathrm{NHIM}$
of the NHIM for $(R = 4.3, p_R = 0)$.
The white dashed square trajectory
is initialized closest to one of the manifolds and stays in the
barrier region for nearly three oscillations in the stable direction of the barrier.
Hence, it crosses a corresponding \ac{DS}
separating reactants from products closest to the \ac{NHIM}.
While the structures illustrated in Fig.~\ref{fig:LiCN_trajectories+state}
were previously seen in a generic model system with a
driven rank-1 saddle,\cite{hern17h,hern18g,hern19a,hern19e}
the results here show that they are realized also in
a specific model of a chemical reaction.
In both cases, the geometry is described by
the stable and unstable manifolds of the barrier, and
the nature of the reactivity is determined by
the associated \ac{DS} attached to the \ac{NHIM}.

\subsection{Dynamics of periodically driven transition states}
\label{sec:results_driven_dynamics}
When periodically driving the LiCN $\rightleftharpoons$ LiNC isomerization
reaction,
the potential
$V\sno{dip}(R, \vartheta, t)$ in Eq.~\eqref{eq:V_dip}
is explicitly time-dependent and
the energy of the system is no longer conserved.
Still,
a two-dimensional \ac{NHIM} exists in the barrier region,
which contains all trajectories that never leave this region,
neither forward, nor backward in time.
In contrast to a static system, however, the \ac{NHIM} becomes
time-dependent.
For periodic driving, such a \ac{NHIM} oscillates with the same frequency
as that of the driving of the barrier
because of the symmetry of the system
with respect to time.

The \ac{PSOS} is a
useful tool for resolving the dynamics of this time-dependent \ac{NHIM}
in a periodically driven system.
Therein, the positions $(R, p_R)$ of several trajectories on the \ac{NHIM}
are marked
after integer multiples of the driving period
for a time much longer than a single period of the external driving.
When propagating trajectories for such a relatively long time, the instability
of the dynamics in the \ac{NHIM} can become problematic
as errors produced by any numerical propagator are exponentially increasing.
Consequently, trajectories initialized as precisely as possible in the \ac{NHIM}
nevertheless
``fall down'' from the barrier to either the reactant or the product side.
This problem can be addressed through the use of a stabilized propagator,\cite{hern20d,hern19e}
which successively projects unstable trajectories back into the
\ac{NHIM} after an appropriately chosen time step using the \ac{BCM}.

\begin{figure}
\includegraphics[width=\columnwidth]{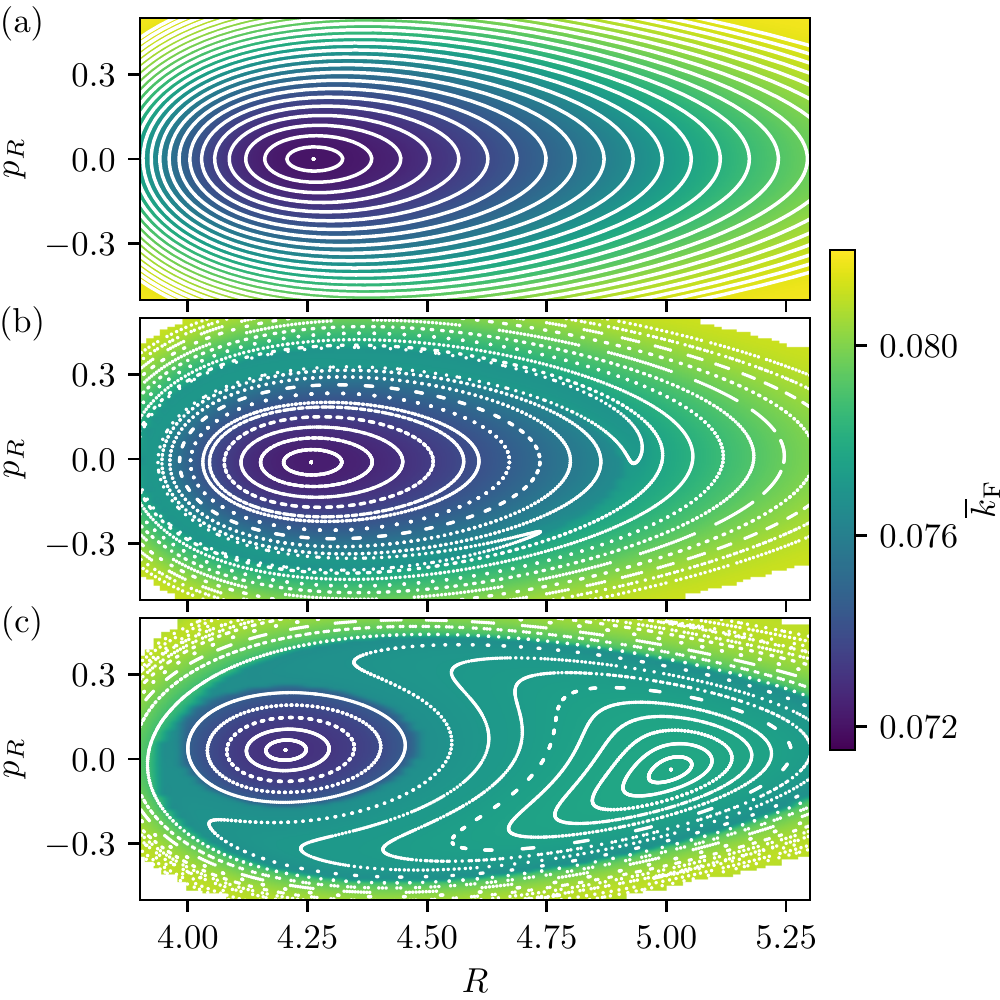}
\caption{%
The stable structures obtained from the dynamics on the \ac{NHIM}
are compared to Floquet rates $\overline{k}\sno{F}$ for the decay
of trajectories from the \ac{NHIM}
for the static (a) and the driven LiCN isomerization reaction with
$E_0 = 0.01$ and $\omega = 0.01\,\pi$ in (b) and $\omega = 0.02\,\pi$ in (c), respectively.
The white curves
in all three panels are the \acp{PSOS} obtained
from periodic orbits in (a), and from the
propagation of trajectories for $400$ periods of the external driving
while marking their instantaneous position $(R, p_R)$
after each period in (b) and (c).
The color encoding shows the mean Floquet rates
obtained for $100 \times 100$ stabilized trajectories which are initialized
on an equidistant grid for the ranges of $R \in [3.8, 5.3]$ and
$p_R \in [-0.5, 0.5]$ on the \ac{NHIM}.
Its contours (not shown but indicated through the color variation)
coincide with the \ac{PSOS}.
All trajectories are stabilized to the \ac{NHIM} using the \ac{BCM} with a tolerance of $10^{-12}$ and
the colors are interpolated using bicubic splines.
}\label{fig:LiCN_psos_together}
\end{figure}

In the static case according to Eq.~\eqref{LiCN_Hamiltonian_rotationless}, the energy is conserved, so each trajectory in
Fig.~\ref{fig:LiCN_psos_together}~(a) is periodic.
The central fixed point corresponds to the trajectory resting at
the saddle point of the barrier---\emph{viz.}\ at $R = 4.2626$ and $\vartheta = 0.2800\,\pi$.
The frequency of oscillations in the stable direction of the static barrier
decreases monotonously from about $\omega\sno{orth} = 0.043\,\pi$
for the trajectories close to the fixed point
to, \eg, about $\omega\sno{orth} = 0.035\,\pi$ for a trajectory with energy $E = -0.19$.
The driving frequencies, chosen below,
range between 2 to 4 times slower than the
oscillation frequencies of the orthogonal modes
and thus provide a non-trivial perturbation to the system.
The \acp{PSOS} at these driving frequencies
indeed show significant changes
compared to the
static case
as shown in Fig.~\ref{fig:LiCN_psos_together}.

For the driven systems,
the stabilized trajectories are initiated
on the \ac{NHIM} at $t_0 = 0$.
They are propagated for $400$ periods
of the external driving
with amplitude $E_0 = 0.01$ and frequency $\omega = 0.01\,\pi$.
Specifically, the values are chosen to
correspond to an electromagnetic field with frequency \SI{4.85}{\THz}
(wavelength $\lambda = \SI{61.9}{\um}$)%
\footnote{These values have been updated to reflect
a correction due to the unit conversion noted in 
footnote~\citenum{Note1}.}
and amplitude $E_0 = \SI{5.14e7}{\V\per\cm}$.
After each period of $T = 200$, the instantaneous position $(R, p_R)$ of each
trajectory in the \ac{NHIM} is marked.
The stabilization of each trajectory into the \ac{NHIM} was performed using
the \ac{BCM} with an error tolerance of $10^{-12}$.
Finally, we observe that
the \acp{PSOS} for the driving cases look quite different from the static case
because the driven \ac{NHIM} is time-dependent.

The driven trajectories are generally no longer periodic and energy is not conserved,
leading to a more complex geometric structure.
Nevertheless, at first glance, the structure of the trajectories of the driven barrier
in Fig.~\ref{fig:LiCN_psos_together}~(b)
looks very similar to those of the non-driven barrier
in Fig.~\ref{fig:LiCN_psos_together}~(a)
having a central fixed point and tori (trajectories in the non-driven case) around it.
On the right-hand side of Fig.~\ref{fig:LiCN_psos_together}~(b) at about $R = 4.8$
and $p_R = -0.2$, an unstable fixed point is visible,
interruption the regular arrangement of the tori.
According to the Poincar\'{e}-Birkhoff theorem,\cite{wimberger2014nonlinear}
only an even number of fixed points may occur in a perturbed system (if regarded not precisely at the
bifurcation values),
and consequently, the appearance of the unstable fixed point is accompanied
by the emergence of a stable fixed point.
The latter is located
approximately at $R = 3.9$ and $p_R = 0$ and is not directly visible here.

To magnify the influence of the moving barrier, the \ac{PSOS}
in Fig.~\ref{fig:LiCN_psos_together}~(c)
is obtained in the
same way as in Fig.~\ref{fig:LiCN_psos_together}~(b) but with twice
the frequency of the external driving $\omega = 0.02\,\pi$.
Physically, this
corresponds to an electromagnetic field with frequency \SI{9.70}{\THz}
(wavelength $\lambda = \SI{30.9}{\um}$).%
\cite{Note2}
Now, the external driving has a strong impact on the dynamics of
the trajectories on the \ac{NHIM}.
An additional unstable and
an additional stable fixed point are clearly visible.
Both stable fixed points belong to separate period-1 trajectories,
that are encircled by many quasi-periodic trajectories on various tori.
These structures,
not seen in the non-driven barrier and only partially seen with weak driving,
arise solely due to the external driving.

\begin{figure}
\includegraphics[width=\columnwidth]{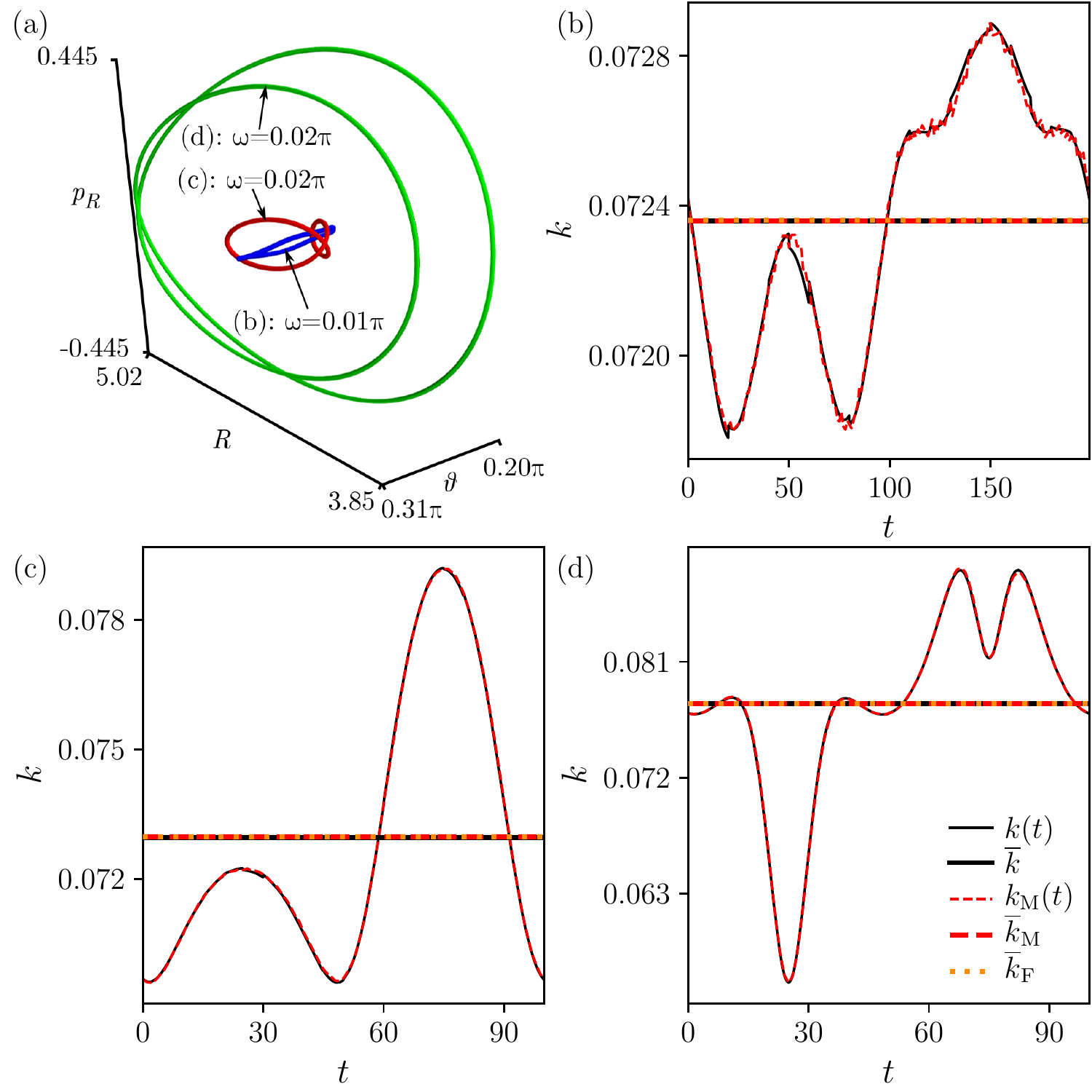}
\caption{%
(a) Representation $(R, \vartheta, p_R)$
of period-1 trajectories on the NHIM of the driven LiCN isomerization
reaction.
The inner trajectory corresponds to the central elliptic fixed
point in Fig.~\ref{fig:LiCN_psos_together}~(b),
the intermediate and outer double-loop trajectories to the elliptic
fixed points in
Figs.~\ref{fig:LiCN_psos_together}~(c) and \ref{fig:LiCN_psos_together}~(d)
at positions $R \approx 4.2$
and $R \approx 5.0$, respectively.
Parts~(b)-(d) show the instantaneous ensemble rate $k(t)$ (thin black line),
the time-averaged mean ensemble rate $\overline{k}$ (thick black horizontal line),
the mean Floquet rate $\overline{k}\sno{F}$ (orange dotted horizontal line),
the instantaneous manifold rate $k\sno{M}(t)$
(thin red dashed line),
and the time-averaged
mean manifold rate $\overline{k}\sno{M}$ (thick red horizontal dashed line).
The methods for obtaining these various rates are described in the
supplementary material.
}\label{fig:3D_trajectories_and_rates}
\end{figure}

A three-dimensional representation $(R, \vartheta, p_R)$
of the three periodic trajectories corresponding to the
stable fixed points clearly shown in Fig.~\ref{fig:LiCN_psos_together}~(b)
and (c) is given in Fig.~\ref{fig:3D_trajectories_and_rates}~(a).
Here, the inner trajectory, corresponding to the visible stable fixed point
in Fig.~\ref{fig:LiCN_psos_together}~(b) at initial position
$R = 4.2579$ and $p_R = -0.0117$ shows just very little
movement in the stable direction of the barrier.
However, the trajectory in between, corresponding to the stable fixed point
at $R = 4.2040$ and $p_R = 0.0313$
and especially the outer trajectory, corresponding to the stable fixed point
at $R = 5.0096$ and $p_R = -0.0378$
of Fig.~\ref{fig:LiCN_psos_together}~(c)
are subject to significant movement in the stable direction of the barrier.
Note, that these two trajectories show two oscillations in
direction of the orthogonal mode of the barrier
although they are both period-1 trajectories
with respect to the external driving.

\subsection{Instantaneous decay near periodic trajectories on the driven NHIM}

Using the methods
introduced in Ref.~\onlinecite{hern19e} and discussed as supplementary material,
the instantaneous
reactant decay rates associated with trajectories on the driven \ac{NHIM}
can be obtained.
First, we analyze
the three period-1 trajectories displayed in Fig.~\ref{fig:3D_trajectories_and_rates}~(a).
The inner
trajectory---shown in blue (in color) on the (a) panel---is
obtained for a driving frequency of $\omega=0.01\,\pi$, while
the two outer
trajectories---shown in blue and green (in color) on the (a) panel---are
obtained for a system with
a larger driving frequency of $\omega=0.02\,\pi$, as labeled.

Figure~\ref{fig:3D_trajectories_and_rates}~(b) presents
the instantaneous decay rates associated with trajectories close to the
inner period-1 trajectory
with an initial point
$R = 4.2579$ and $p_R = -0.0117$ at time $t=0$
on the \ac{NHIM}
of a driven barrier with $E_0 = 0.01$ and $\omega = 0.01\,\pi$.
To obtain decay rates with the ensemble method,
the trajectory is divided into $20$ segments.
For each segment, an ensemble of $200$ reactive trajectories
is initialized close to the \ac{NHIM} with a distance of $\Delta \vartheta = 10^{-3}$.
Patching together these individual segments yields the thin black line in
Fig.~\ref{fig:3D_trajectories_and_rates}~(b). The corresponding mean ensemble decay rate
is obtained by averaging over a full period of the external driving,
yielding $\overline{k} = 0.0724$ and displayed as a
thick solid vertical line.
This result can be verified using the Floquet method,
see supplementary material,
which yields a mean Floquet rate
of $\overline{k}\sno{F} = 0.0724$ for this trajectory.
This mean Floquet rate,
shown as an orange dotted vertical line in Fig.~\ref{fig:3D_trajectories_and_rates}~(b),
is in perfect agreement with the mean ensemble rate.
Using the \ac{LMA}, see supplementary material,
a third verification
of these decay rates can be obtained. The thin red dashed line
in Fig.~\ref{fig:3D_trajectories_and_rates}~(b) marks the instantaneous manifold rate
$k\sno{M}(t)$ and the thick red dashed vertical line marks
the mean manifold rate $\overline{k}\sno{M} = 0.0724$,
averaged over a full period of the external driving.
Both lines fit the results of the other two methods.

The same procedure used to obtain Fig.~\ref{fig:3D_trajectories_and_rates}~(b) is repeated
for the two remaining trajectories of Fig.~\ref{fig:3D_trajectories_and_rates}~(a).
Figures~\ref{fig:3D_trajectories_and_rates}~(c)
and \ref{fig:3D_trajectories_and_rates}~(d)
present the results
for the intermediate trajectory
(labeled c, and red in color),
and
the outer trajectory
(labeled d, and green in color)
in Fig.~\ref{fig:3D_trajectories_and_rates}~(a), respectively.
In both cases,
the instantaneous decay rates obtained via the \ac{LMA} correspond perfectly
to the instantaneous decay rates of the ensemble method,
and their mean rates are also in perfect agreement with the obtained Floquet rates.
Hence, $\overline{k} = \overline{k}\sno{M} = \overline{k}\sno{F} = 0.0730$
for the intermediate trajectory,
and $\overline{k} = \overline{k}\sno{M} = \overline{k}\sno{F} = 0.0778$
for the outer trajectory.

The decay of the reactant population close to
the three different period-1 trajectories
in Figs.~\ref{fig:3D_trajectories_and_rates}~(b)-(d)
is represented by very different
instantaneous rates, and consequently also mean rates.
In addition, the amount of variation of the instantaneous rates
varies strongly for each trajectory.
For the inner trajectory
that according to Fig.~\ref{fig:3D_trajectories_and_rates}~(a)
has the smallest movement in the orthogonal mode,
the relative change of the instantaneous rate,
$\Delta k \approx 10^{-3}$, is very small.
Consequently, the effect of the external driving is barely noticeable.
The intermediate trajectory
of Fig.~\ref{fig:3D_trajectories_and_rates}~(a)
has significantly more motion
in the stable direction of the barrier.
The relative change, $\Delta k \approx 10^{-2}$, of the
instantaneous rate is considerably larger as seen in
Fig.~\ref{fig:3D_trajectories_and_rates}~(c).
The outer trajectory of Fig.~\ref{fig:3D_trajectories_and_rates}~(a)
has by far the largest movement in the stable direction of the barrier.
The relative change, $\Delta k \approx 0.5\times 10^{-1}$,
in the instantaneous rates in Fig.~\ref{fig:3D_trajectories_and_rates}~(d)
is also the largest.
Thus, we can conclude that
the influence of the external driving is high
if a trajectory has significant motion in the direction
of the orthogonal modes.
Further evidence for this effect
follows in Sec.~\ref{sec:phase-space_resolved_rates_driven_licn} in the context of
the decay rates associated with the numerous quasi-periodic trajectories
on the driven \ac{NHIM}.

\subsection{Phase-space resolved decay rates}
\label{sec:phase-space_resolved_rates_driven_licn}
We now obtain the phase-space resolved decay rates of
the reactant population close to
arbitrary trajectories on the \ac{NHIM} of
the periodically driven LiCN $\rightleftharpoons$ LiNC
isomerization reaction.
As seen above, all three methods
to calculate the decay of reactant population close to the \ac{TS}
result in the same values for a specific
period-1 trajectory, and hence we choose only one of these for
the present calculation.
Namely, the Floquet method is employed because it is
relatively easy to implement and computationally fast to evaluate.

In all the cases of Fig.~\ref{fig:LiCN_psos_together},
the Floquet rates are overlayed on top of the \acp{PSOS}.
They are obtained on equidistant grids
using approximately \num{10000} stabilized trajectories,
and they are displayed
through a shaded (or colored in color) encoding.
All trajectories are propagated for a total time up to
$t = \num{10000}$ as needed to converge.
The individual Floquet rates are interpolated using bicubic splines
to smooth the discrete points.
We found that the decay rates of all the trajectories located on
the corresponding regular tori are indeed the same.
This is expected since any quasi-periodic trajectory on such a torus
in general covers it in full if propagated for long enough.
Thus, a unique decay rate is associated with each torus.
This observation is true for the static system using a simpler
argument.
Namely, the Floquet rate reduces to a property of any of the periodic
trajectories on the static \ac{NHIM} because the tori is necessarily
periodic.

When comparing the obtained mean Floquet rates of the driven system
with $E_0 = 0.01$ and $\omega = 0.01\,\pi$
in Fig.~\ref{fig:LiCN_psos_together}~(b) to the mean Floquet rates of the
static system according to Fig.~\ref{fig:LiCN_psos_together}~(a),
the influence of the external driving
is small.
In Fig.~\ref{fig:LiCN_psos_together}~(b),
the mean Floquet rate obtained at the
clearly visible elliptic fixed point in the center of the
tori corresponds to the rates already obtained in Fig.~\ref{fig:3D_trajectories_and_rates}~(b)
with all three methods
introduced in Ref.~\onlinecite{hern19e} and provided as supplementary material.
In the two cases of
Fig.~\ref{fig:LiCN_psos_together}~(a) and
Fig.~\ref{fig:LiCN_psos_together}~(b),
the mean Floquet rates at the central fixed point
are similar---that is, they
are $\overline{k}\sno{F} = 0.0721$ and $\overline{k}\sno{F} = 0.0724$, respectively.
The primary difference is the emergence of structure
that is barely visible here but more clearly visible in
Fig.~\ref{fig:LiCN_psos_together}~(c) as discussed below.

A further increase in the frequency
of the external driving to $\omega = 0.02\,\pi$
leads to a drastic change in the structure as shown in
Fig.~\ref{fig:LiCN_psos_together}~(c).
Two clearly separated elliptical fixed points are now visible in the \ac{PSOS}
with each encircled by an individual
set of tori dividing the \ac{NHIM} into two regions of very different Floquet rates.
This means that two regions of very different stability emerge on
the periodically driven \ac{NHIM}.
In the surrounding of the fixed point at $R=4.2040$ and $p_R = 0.0313$,
the decay rates of the reactant population are rather small and approximately correspond to
the mean Floquet rate of $\overline{k}\sno{F} = 0.0730$ obtained
for the central period-1 trajectory according to Fig.~\ref{fig:3D_trajectories_and_rates}~(c).
On the other hand,
the mean Floquet rates in a region near the fixed point at
$R = 5.0096$ and $p_R = -0.0378$ are rather large and approximately
correspond to the mean Floquet rate $\overline{k}\sno{F}=0.0778$
of the associated central period-1 trajectory
according to Fig.~\ref{fig:3D_trajectories_and_rates}~(d).

\begin{figure}
\includegraphics[width=0.9\columnwidth]{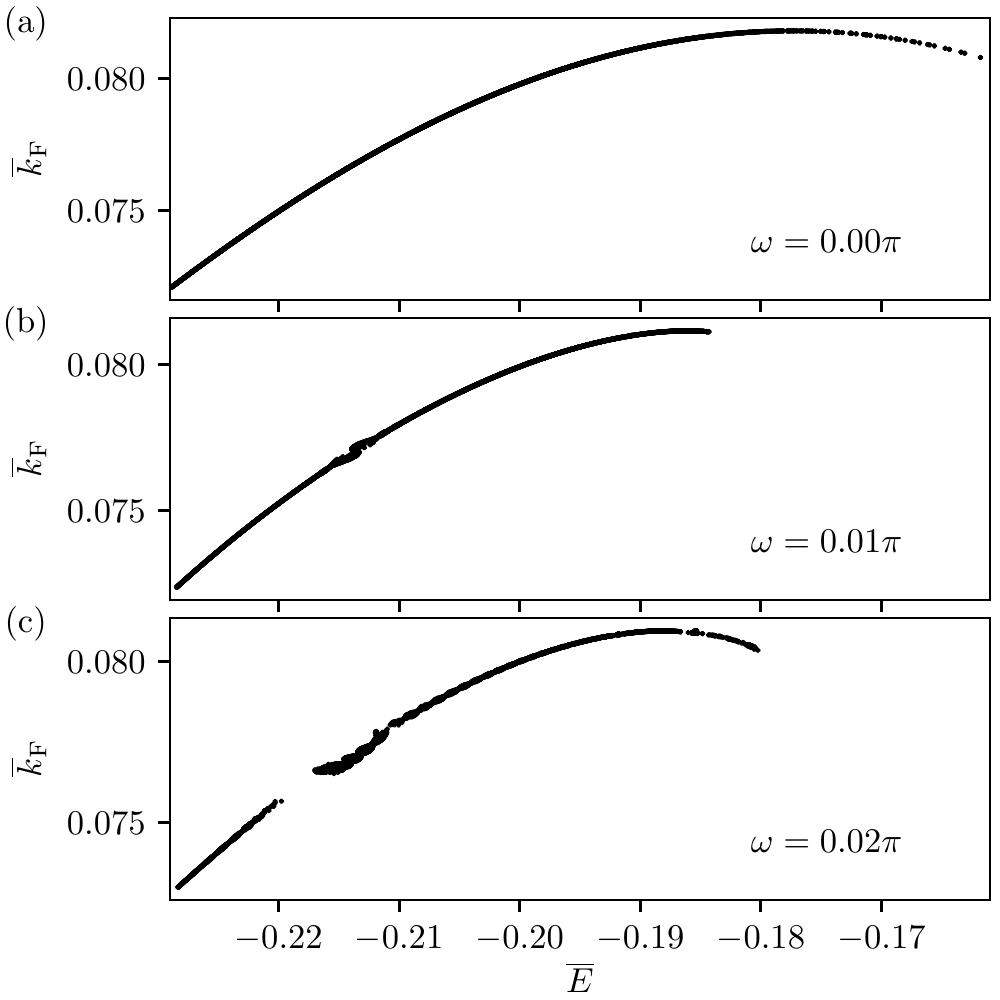}
\caption{%
The quantitative relation of the Floquet rates
$\overline{k}\sno{F}$
to the
mean energies $\overline{E}$ of trajectories on the \ac{NHIM}
obtained at three different driving frequencies $\omega$ equal to
$0.00\,\pi$, $0.01\,\pi$ and $0.02\,\pi$
are shown in panels (a), (b) and (c), respectively.
Each rate corresponds to a torus in
Fig.~\ref{fig:LiCN_psos_together} at the mean energy
$\overline{E}$.
}\label{fig:LiCN_energy_rate_together}
\end{figure}

For a periodically driven system,
the instantaneous energy $E(t)$ of trajectories in the \ac{NHIM} is not conserved.
However,
by averaging the instantaneous and non-conserved
energy of a given trajectory over many oscillations of the
periodic driving, the
mean energy $\overline{E}$ of a specific torus is well defined
and characteristic of the trajectory and its associated initial condition.
The quantitative correspondence
between the mean energy of a trajectory
and the Floquet rate
is revealed in Fig.~\ref{fig:LiCN_energy_rate_together}
by plotting the Floquet rates obtained in
Fig.~\ref{fig:LiCN_psos_together}
with respect to the corresponding (mean) energies of the trajectories.

As a control for the driven cases,
Fig.~\ref{fig:LiCN_energy_rate_together}~(a) shows
the relation between the Floquet rate and the mean energy
for the non-driven system
of Fig.~\ref{fig:LiCN_psos_together}~(a).
In this limit, energy is conserved.
Thus the mean energy $\overline{E}$ corresponds
to the instantaneous energy $E(t)$ of the respective trajectories.
According to Fig.~\ref{fig:LiCN_energy_rate_together}~(a),
the relation between the energy and the Floquet rate in the static case
is not monotonic and has a maximum at
approximately $E=-0.18$.
A further increase of the energy of trajectories
on the \ac{NHIM} goes together with a decrease in the corresponding decay rate
of the reactant population close to the \ac{TS}.
However,
a maximum in these decay rates is associated with a minimum in the stability of the
activated complex close to the \ac{TS}.
According to Fig.~\ref{fig:LiCN_energy_rate_together}~(a),
there seems to be a lower bound for the stability of
this activated complex
in the non-driven LiCN isomerization reaction.

The results for the driven system
in Figs.~\ref{fig:LiCN_energy_rate_together}~(b) for
$\omega = 0.01\,\pi$ and in Fig.~\ref{fig:LiCN_energy_rate_together}~(c)
for $\omega=0.02\,\pi$
correspond to cases in Fig.~\ref{fig:LiCN_psos_together}~(b) and (c), respectively.
Despite the driving, they retain some
similarity with the static case of Fig.~\ref{fig:LiCN_energy_rate_together}~(a).
Again, the relation between the Floquet rate of a trajectory on a given torus
and the associated mean energy is non-monotonic and
there seems to exist a maximum decay rate
for the reactant population associated with trajectories on the \ac{NHIM}.

In contrast to the static case, however,
for the driven barrier case with $\omega=0.02\,\pi$,
a significant gap arises in the curve of
Fig.~\ref{fig:LiCN_energy_rate_together}~(c)
at about $\overline{E}=-0.22$.
This gap corresponds
exactly to the
emergence of tori around distinct fixed points seen in
Fig.~\ref{fig:LiCN_psos_together}~(c).
At the beginning of the
upper edge of the gap in Fig.~\ref{fig:LiCN_energy_rate_together}~(c),
the curve
initially fluctuates before becoming smooth again
with increasing mean energy.
The origin of this behavior is numerical
because the Floquet method for non-periodic trajectories
is defined only in the limit $t\to \infty$ (see supplementary material).
However, trajectories in Fig.~\ref{fig:LiCN_psos_together}~(c)
are necessarily propagated for a finite time,
and this leads to errors in the calculated Floquet rates
which are small and appear as fluctuations.
In other words, close to the boundary
between the two regions of the different decay rates on the driven \ac{NHIM},
the circulation times of quasi-periodic trajectories on the regular tori increases
and longer propagation times are necessary
to obtain a desired accuracy using the Floquet method.
This behavior can also be seen in Fig.~\ref{fig:LiCN_energy_rate_together}~(b)
for the case with a driving frequency of $\omega=0.01\,\pi$.
A second stable fixed point is barely visible
in Fig.~\ref{fig:LiCN_psos_together}~(b).
Consequently, nearly no gap appears in Fig.~\ref{fig:LiCN_energy_rate_together}~(b).
Nevertheless, fluctuations are visible at the energy
of the indicated fixed point on the left-hand side of
Fig.~\ref{fig:LiCN_psos_together}~(b),
and, hence, a gap is just emerging there.

\section{Conclusion and Outlook}
\label{sec:Conclusion}

In this paper, we
study the influence of a periodically oscillating
external field on the LiCN $\rightleftharpoons$ LiNC isomerization reaction.
In the static case, the energy is conserved and
all trajectories on the associated \ac{NHIM} are periodic.
For a periodically driven system,
nearly all trajectories contained in the NHIM are quasi-periodic
and located on stable tori.
The dynamics on the driven \ac{NHIM} has been revealed by means of a \ac{PSOS}.
Depending on the frequency of the external field,
a new set of tori around a second stable fixed point
emerges on the \ac{NHIM}.

The associated mean decay rates of reactant population
in a close neighborhood of the periodic or quasi-periodic trajectories
on the \ac{NHIM} are obtained by various methods.\cite{hern19e}
Both in the static and driven cases,
the decay rates differ significantly across these neighborhoods,
and are approximately prescribed by the mean decay rates associated to
the corresponding central periodic trajectory.
The instantaneous decay rates
for these central trajectories are also sensitive to the
driving.
Specifically, the relation between the mean decay rates associated with
a specific torus and the corresponding mean energy
of trajectories on this torus is
non-trivial.
There are two clearly distinguished regions
associated with different mean decay rates
of the reactant population close to the \ac{TS}.
They are approximately prescribed by period-1 trajectories
identified at the center of the corresponding tori of the \ac{PSOS}.
They are consequently not continuously connected
and a significant gap between the corresponding decay rate
emerges.
The regularity of the structure of the \ac{PSOS} for the trajectories
near the \ac{NHIM}
is not chaotic as one might have expected given the
recently observed chaotic sea in the global phase space.\cite{Revuelta19}
This points to the simplification arising from considering decay rates
in the local neighborhood of the \ac{TS}.
Future work to extend these decay rates to the global rates thus needs
to consider not just the possibility of additional barrier regions but
also the complex dynamics that can arise from chaotic regions.

Thus the influence of the periodic external driving
on the LiCN $\rightleftharpoons$ LiNC isomerization reaction
is large
when compared to the static problem.
Indeed, the emergence of different regions of the reactant population decay
for the driven LiCN $\rightleftharpoons$ LiNC isomerization reaction
as well as the existence of gaps in both energy and decay rate
between these separate regions are effects solely invoked
by the external driving.

The possible effects from our approximate simplification
of the LiCN $\rightleftharpoons$ LiNC isomerization reaction
to a body-fixed axes may need to be addressed.
This would allow the molecules to rotate
with respect to the external field
and increase the dimensionality of the coordinate space of the problem
beyond the two-dimensional case considered here.
However, there is precedent from the earlier work of
Ref.~\onlinecite{murgida2015quantum} that the reaction is much faster
than the rotation and hence such effects are small.
One could also
include a Langevin-type description of the
dynamics---according to, \eg, Ref.~\onlinecite{hern16c}---to
address the effects on the decay rates from a solvent
as represented by the inclusion of both noise and friction.
Alternatively, the effects on
the decay rates from an all-atom solvent
can be uncovered by inferring the stability of the
\ac{NHIM} from molecular dynamics trajectories.
For example, the position of reactive trajectories across the \ac{DS}
can be tracked simulations of LiCN
in an argon bath such as those of
Refs.~\onlinecite{hern08g,hern12e,hern14j,hern16c}.
Applying an appropriate external driving could
shift these reaction positions to different regions of the \ac{NHIM}
and increases the reaction rate.
From the results of the current work, we would expect that
there would once again emerge different regions in the
\ac{PSOS} and the corresponding reactant decay rates
which should be interpretable as reactive channels.
When applying an appropriate external field
a second and faster reactive channel
with higher decay rates
opens up and might serve to
an increase of the reaction rate.

\section*{Supplementary material}

In the supplementary material, we provide a detailed
description of the methods used in the primary text.
We summarize the details for the implementation of three approaches in
determining the decay rates: Floquet analysis, ensemble method, and the
local manifold analysis.
We also present a technical derivation of the last of these methods.


\begin{acknowledgments}
    The German portion of this collaborative work was supported
    by Deutsche Forschungsgemeinschaft (DFG) through Grant
    No.~MA1639/14-1.
    RH's contribution to this work was supported by the National Science
    Foundation (NSF) through Grant No.~CHE-1700749.
    M.F.\ is grateful for support from the Landesgraduiertenf\"orderung of
    the Land Baden-W\"urttemberg.
    This collaboration has also benefited from support by the European
    Union's Horizon 2020 Research and Innovation Program under the Marie
    Sklodowska-Curie Grant Agreement No.~734557.
\end{acknowledgments}

\section*{Data Availability}

The data that support the findings of this study are available from the corresponding author upon reasonable request.


\section*{References}
\bibliographystyle{apsrev4-1-forcedoi}
\bibliography{paper-q21}

\begin{thebibliography}{71}%
\makeatletter
\providecommand \@ifxundefined [1]{%
 \@ifx{#1\undefined}
}%
\providecommand \@ifnum [1]{%
 \ifnum #1\expandafter \@firstoftwo
 \else \expandafter \@secondoftwo
 \fi
}%
\providecommand \@ifx [1]{%
 \ifx #1\expandafter \@firstoftwo
 \else \expandafter \@secondoftwo
 \fi
}%
\providecommand \natexlab [1]{#1}%
\providecommand \enquote  [1]{``#1''}%
\providecommand \bibnamefont  [1]{#1}%
\providecommand \bibfnamefont [1]{#1}%
\providecommand \citenamefont [1]{#1}%
\providecommand \href@noop [0]{\@secondoftwo}%
\providecommand \href [0]{\begingroup \@sanitize@url \@href}%
\providecommand \@href[1]{\@@startlink{#1}\@@href}%
\providecommand \@@href[1]{\endgroup#1\@@endlink}%
\providecommand \@sanitize@url [0]{\catcode `\\12\catcode `\$12\catcode
  `\&12\catcode `\#12\catcode `\^12\catcode `\_12\catcode `\%12\relax}%
\providecommand \@@startlink[1]{}%
\providecommand \@@endlink[0]{}%
\providecommand \url  [0]{\begingroup\@sanitize@url \@url }%
\providecommand \@url [1]{\endgroup\@href {#1}{\urlprefix }}%
\providecommand \urlprefix  [0]{URL }%
\providecommand \Eprint [0]{\href }%
\providecommand \doibase [0]{http://dx.doi.org/}%
\providecommand \selectlanguage [0]{\@gobble}%
\providecommand \bibinfo  [0]{\@secondoftwo}%
\providecommand \bibfield  [0]{\@secondoftwo}%
\providecommand \translation [1]{[#1]}%
\providecommand \BibitemOpen [0]{}%
\providecommand \bibitemStop [0]{}%
\providecommand \bibitemNoStop [0]{.\EOS\space}%
\providecommand \EOS [0]{\spacefactor3000\relax}%
\providecommand \BibitemShut  [1]{\csname bibitem#1\endcsname}%
\let\auto@bib@innerbib\@empty
\bibitem [{\citenamefont {Eyring}(1935)}]{eyring35}%
  \BibitemOpen
  \bibfield  {author} {\bibinfo {author} {\bibfnamefont {H.}~\bibnamefont
  {Eyring}},\ }\href {\doibase 10.1063/1.1749604} {\bibfield  {journal}
  {\bibinfo  {journal} {J. Chem. Phys.}\ }\textbf {\bibinfo {volume} {3}},\
  \bibinfo {pages} {107} (\bibinfo {year} {1935}),\
  doi:10.1063/1.1749604}\BibitemShut {NoStop}%
\bibitem [{\citenamefont {Wigner}(1937)}]{wigner37}%
  \BibitemOpen
  \bibfield  {author} {\bibinfo {author} {\bibfnamefont {E.~P.}\ \bibnamefont
  {Wigner}},\ }\href {\doibase 10.1063/1.1750107} {\bibfield  {journal}
  {\bibinfo  {journal} {J. Chem. Phys.}\ }\textbf {\bibinfo {volume} {5}},\
  \bibinfo {pages} {720} (\bibinfo {year} {1937}),\
  doi:10.1063/1.1750107}\BibitemShut {NoStop}%
\bibitem [{\citenamefont {Pollak}\ \emph {et~al.}(1980)\citenamefont {Pollak},
  \citenamefont {Child},\ and\ \citenamefont {Pechukas}}]{pollak80}%
  \BibitemOpen
  \bibfield  {author} {\bibinfo {author} {\bibfnamefont {E.}~\bibnamefont
  {Pollak}}, \bibinfo {author} {\bibfnamefont {M.~S.}\ \bibnamefont {Child}}, \
  and\ \bibinfo {author} {\bibfnamefont {P.}~\bibnamefont {Pechukas}},\ }\href
  {\doibase 10.1063/1.439276} {\bibfield  {journal} {\bibinfo  {journal} {J.
  Chem. Phys.}\ }\textbf {\bibinfo {volume} {72}},\ \bibinfo {pages} {1669}
  (\bibinfo {year} {1980}),\ doi:10.1063/1.439276}\BibitemShut {NoStop}%
\bibitem [{\citenamefont {Pechukas}(1981)}]{pech81}%
  \BibitemOpen
  \bibfield  {author} {\bibinfo {author} {\bibfnamefont {P.}~\bibnamefont
  {Pechukas}},\ }\href {\doibase 10.1146/annurev.pc.32.100181.001111}
  {\bibfield  {journal} {\bibinfo  {journal} {Annu. Rev. Phys. Chem.}\ }\textbf
  {\bibinfo {volume} {32}},\ \bibinfo {pages} {159} (\bibinfo {year} {1981}),\
  doi:10.1146/annurev.pc.32.100181.001111}\BibitemShut {NoStop}%
\bibitem [{\citenamefont {Truhlar}\ \emph {et~al.}(1996)\citenamefont
  {Truhlar}, \citenamefont {Garrett},\ and\ \citenamefont
  {Klippenstein}}]{truh96}%
  \BibitemOpen
  \bibfield  {author} {\bibinfo {author} {\bibfnamefont {D.~G.}\ \bibnamefont
  {Truhlar}}, \bibinfo {author} {\bibfnamefont {B.~C.}\ \bibnamefont
  {Garrett}}, \ and\ \bibinfo {author} {\bibfnamefont {S.~J.}\ \bibnamefont
  {Klippenstein}},\ }\href {\doibase 10.1021/jp953748q} {\bibfield  {journal}
  {\bibinfo  {journal} {J. Phys. Chem.}\ }\textbf {\bibinfo {volume} {100}},\
  \bibinfo {pages} {12771} (\bibinfo {year} {1996}),\
  doi:10.1021/jp953748q}\BibitemShut {NoStop}%
\bibitem [{\citenamefont {Carpenter}(2005)}]{Carpenter2005a}%
  \BibitemOpen
  \bibfield  {author} {\bibinfo {author} {\bibfnamefont {B.~K.}\ \bibnamefont
  {Carpenter}},\ }\enquote {\bibinfo {title} {Potential energy surfaces and
  reaction dynamics},}\ in\ \href {\doibase 10.1002/0471721492.ch21} {\emph
  {\bibinfo {booktitle} {Reactive Intermediate Chemistry}}}\ (\bibinfo
  {publisher} {John Wiley \& Sons, Ltd},\ \bibinfo {year} {2005})\
  Chap.~\bibinfo {chapter} {21}, pp.\ \bibinfo {pages} {925--960}\BibitemShut
  {NoStop}%
\bibitem [{\citenamefont {Mullen}\ \emph {et~al.}(2014)\citenamefont {Mullen},
  \citenamefont {Shea},\ and\ \citenamefont {Peters}}]{peters14a}%
  \BibitemOpen
  \bibfield  {author} {\bibinfo {author} {\bibfnamefont {R.~G.}\ \bibnamefont
  {Mullen}}, \bibinfo {author} {\bibfnamefont {J.-E.}\ \bibnamefont {Shea}}, \
  and\ \bibinfo {author} {\bibfnamefont {B.}~\bibnamefont {Peters}},\ }\href
  {\doibase 10.1063/1.4862504} {\bibfield  {journal} {\bibinfo  {journal} {J.
  Chem. Phys.}\ }\textbf {\bibinfo {volume} {140}},\ \bibinfo {pages} {041104}
  (\bibinfo {year} {2014}),\ doi:10.1063/1.4862504}\BibitemShut {NoStop}%
\bibitem [{\citenamefont {Arrhenius}(1889)}]{Arrhenius1889}%
  \BibitemOpen
  \bibfield  {author} {\bibinfo {author} {\bibfnamefont {S.}~\bibnamefont
  {Arrhenius}},\ }\href@noop {} {\bibfield  {journal} {\bibinfo  {journal} {Z.
  Phys. Chem. (Leipzig)}\ }\textbf {\bibinfo {volume} {4}},\ \bibinfo {pages}
  {226} (\bibinfo {year} {1889})},\ \bibinfo {note} {translated and published
  in Margaret H. Back and Keith J. Laidler, eds., Selected Readings in Chemical
  Kinetics (Oxford: Pergamon, 1967)}\BibitemShut {NoStop}%
\bibitem [{\citenamefont {Laidler}(1984)}]{Laidler1984}%
  \BibitemOpen
  \bibfield  {author} {\bibinfo {author} {\bibfnamefont {K.~J.}\ \bibnamefont
  {Laidler}},\ }\href {\doibase 10.1021/ed061p494} {\bibfield  {journal}
  {\bibinfo  {journal} {J. Chem. Educ.}\ }\textbf {\bibinfo {volume} {61}},\
  \bibinfo {pages} {494} (\bibinfo {year} {1984}),\
  doi:10.1021/ed061p494}\BibitemShut {NoStop}%
\bibitem [{\citenamefont {Pollak}(1990)}]{pollak90a}%
  \BibitemOpen
  \bibfield  {author} {\bibinfo {author} {\bibfnamefont {E.}~\bibnamefont
  {Pollak}},\ }\href {\doibase 10.1063/1.459175} {\bibfield  {journal}
  {\bibinfo  {journal} {J. Chem. Phys.}\ }\textbf {\bibinfo {volume} {93}},\
  \bibinfo {pages} {1116} (\bibinfo {year} {1990}),\
  doi:10.1063/1.459175}\BibitemShut {NoStop}%
\bibitem [{\citenamefont {Pollak}(1996)}]{pollak96}%
  \BibitemOpen
  \bibfield  {author} {\bibinfo {author} {\bibfnamefont {E.}~\bibnamefont
  {Pollak}},\ }in\ \href@noop {} {\emph {\bibinfo {booktitle} {Dynamics of
  Molecules and Chemical Reactions}}},\ \bibinfo {editor} {edited by\ \bibinfo
  {editor} {\bibfnamefont {R.~E.}\ \bibnamefont {Wyatt}}\ and\ \bibinfo
  {editor} {\bibfnamefont {J.}~\bibnamefont {Zhang}}}\ (\bibinfo  {publisher}
  {Marcel Dekker},\ \bibinfo {address} {New York},\ \bibinfo {year} {1996})\
  pp.\ \bibinfo {pages} {617--669}\BibitemShut {NoStop}%
\bibitem [{\citenamefont {Uzer}\ \emph {et~al.}(2002)\citenamefont {Uzer},
  \citenamefont {Jaff\'e}, \citenamefont {Palaci{\'a}n}, \citenamefont
  {Yanguas},\ and\ \citenamefont {Wiggins}}]{uzer02}%
  \BibitemOpen
  \bibfield  {author} {\bibinfo {author} {\bibfnamefont {T.}~\bibnamefont
  {Uzer}}, \bibinfo {author} {\bibfnamefont {C.}~\bibnamefont {Jaff\'e}},
  \bibinfo {author} {\bibfnamefont {J.}~\bibnamefont {Palaci{\'a}n}}, \bibinfo
  {author} {\bibfnamefont {P.}~\bibnamefont {Yanguas}}, \ and\ \bibinfo
  {author} {\bibfnamefont {S.}~\bibnamefont {Wiggins}},\ }\href {\doibase
  10.1088/0951-7715/15/4/301} {\bibfield  {journal} {\bibinfo  {journal}
  {Nonlinearity}\ }\textbf {\bibinfo {volume} {15}},\ \bibinfo {pages} {957}
  (\bibinfo {year} {2002}),\ doi:10.1088/0951-7715/15/4/301}\BibitemShut
  {NoStop}%
\bibitem [{\citenamefont {Komatsuzaki}\ and\ \citenamefont
  {Berry}(2001)}]{KomatsuzakiBerry01a}%
  \BibitemOpen
  \bibfield  {author} {\bibinfo {author} {\bibfnamefont {T.}~\bibnamefont
  {Komatsuzaki}}\ and\ \bibinfo {author} {\bibfnamefont {R.~S.}\ \bibnamefont
  {Berry}},\ }\href {\doibase 10.1073/pnas.131627698} {\bibfield  {journal}
  {\bibinfo  {journal} {Proc. Natl. Acad. Sci. U.S.A.}\ }\textbf {\bibinfo
  {volume} {98}},\ \bibinfo {pages} {7666} (\bibinfo {year} {2001}),\
  doi:10.1073/pnas.131627698}\BibitemShut {NoStop}%
\bibitem [{\citenamefont {Komatsuzaki}\ and\ \citenamefont
  {Berry}(2002)}]{KomatsuzakiBerry02}%
  \BibitemOpen
  \bibfield  {author} {\bibinfo {author} {\bibfnamefont {T.}~\bibnamefont
  {Komatsuzaki}}\ and\ \bibinfo {author} {\bibfnamefont {R.~S.}\ \bibnamefont
  {Berry}},\ }\href {\doibase 10.1002/0471231509.ch2} {\bibfield  {journal}
  {\bibinfo  {journal} {Adv. Chem. Phys.}\ }\textbf {\bibinfo {volume} {123}},\
  \bibinfo {pages} {79} (\bibinfo {year} {2002}),\
  doi:10.1002/0471231509.ch2}\BibitemShut {NoStop}%
\bibitem [{\citenamefont {Bartsch}\ \emph
  {et~al.}(2005{\natexlab{a}})\citenamefont {Bartsch}, \citenamefont
  {Hernandez},\ and\ \citenamefont {Uzer}}]{dawn05a}%
  \BibitemOpen
  \bibfield  {author} {\bibinfo {author} {\bibfnamefont {T.}~\bibnamefont
  {Bartsch}}, \bibinfo {author} {\bibfnamefont {R.}~\bibnamefont {Hernandez}},
  \ and\ \bibinfo {author} {\bibfnamefont {T.}~\bibnamefont {Uzer}},\ }\href
  {\doibase 10.1103/PhysRevLett.95.058301} {\bibfield  {journal} {\bibinfo
  {journal} {Phys. Rev. Lett.}\ }\textbf {\bibinfo {volume} {95}},\ \bibinfo
  {pages} {058301} (\bibinfo {year} {2005}{\natexlab{a}}),\
  doi:10.1103/PhysRevLett.95.058301}\BibitemShut {NoStop}%
\bibitem [{\citenamefont {Pollak}\ and\ \citenamefont
  {Talkner}(2005)}]{pollak05a}%
  \BibitemOpen
  \bibfield  {author} {\bibinfo {author} {\bibfnamefont {E.}~\bibnamefont
  {Pollak}}\ and\ \bibinfo {author} {\bibfnamefont {P.}~\bibnamefont
  {Talkner}},\ }\href {\doibase 10.1063/1.1858782} {\bibfield  {journal}
  {\bibinfo  {journal} {Chaos}\ }\textbf {\bibinfo {volume} {15}},\ \bibinfo
  {pages} {026116} (\bibinfo {year} {2005}),\
  doi:10.1063/1.1858782}\BibitemShut {NoStop}%
\bibitem [{\citenamefont {Bartsch}\ \emph {et~al.}(2008)\citenamefont
  {Bartsch}, \citenamefont {Moix}, \citenamefont {Hernandez}, \citenamefont
  {Kawai},\ and\ \citenamefont {Uzer}}]{hern08d}%
  \BibitemOpen
  \bibfield  {author} {\bibinfo {author} {\bibfnamefont {T.}~\bibnamefont
  {Bartsch}}, \bibinfo {author} {\bibfnamefont {J.~M.}\ \bibnamefont {Moix}},
  \bibinfo {author} {\bibfnamefont {R.}~\bibnamefont {Hernandez}}, \bibinfo
  {author} {\bibfnamefont {S.}~\bibnamefont {Kawai}}, \ and\ \bibinfo {author}
  {\bibfnamefont {T.}~\bibnamefont {Uzer}},\ }\href {\doibase
  10.1002/9780470371572.ch4} {\bibfield  {journal} {\bibinfo  {journal} {Adv.
  Chem. Phys.}\ }\textbf {\bibinfo {volume} {140}},\ \bibinfo {pages} {191}
  (\bibinfo {year} {2008}),\ doi:10.1002/9780470371572.ch4}\BibitemShut
  {NoStop}%
\bibitem [{\citenamefont {Hernandez}\ \emph {et~al.}(2010)\citenamefont
  {Hernandez}, \citenamefont {Bartsch},\ and\ \citenamefont {Uzer}}]{hern10a}%
  \BibitemOpen
  \bibfield  {author} {\bibinfo {author} {\bibfnamefont {R.}~\bibnamefont
  {Hernandez}}, \bibinfo {author} {\bibfnamefont {T.}~\bibnamefont {Bartsch}},
  \ and\ \bibinfo {author} {\bibfnamefont {T.}~\bibnamefont {Uzer}},\ }\href
  {\doibase 10.1016/j.chemphys.2010.01.016} {\bibfield  {journal} {\bibinfo
  {journal} {Chem. Phys.}\ }\textbf {\bibinfo {volume} {370}},\ \bibinfo
  {pages} {270} (\bibinfo {year} {2010}),\
  doi:10.1016/j.chemphys.2010.01.016}\BibitemShut {NoStop}%
\bibitem [{\citenamefont {Waalkens}\ \emph {et~al.}(2008)\citenamefont
  {Waalkens}, \citenamefont {Schubert},\ and\ \citenamefont
  {Wiggins}}]{Waalkens2008}%
  \BibitemOpen
  \bibfield  {author} {\bibinfo {author} {\bibfnamefont {H.}~\bibnamefont
  {Waalkens}}, \bibinfo {author} {\bibfnamefont {R.}~\bibnamefont {Schubert}},
  \ and\ \bibinfo {author} {\bibfnamefont {S.}~\bibnamefont {Wiggins}},\ }\href
  {\doibase 10.1088/0951-7715/21/1/R01} {\bibfield  {journal} {\bibinfo
  {journal} {Nonlinearity}\ }\textbf {\bibinfo {volume} {21}},\ \bibinfo
  {pages} {R1} (\bibinfo {year} {2008}),\
  doi:10.1088/0951-7715/21/1/R01}\BibitemShut {NoStop}%
\bibitem [{\citenamefont {Wiggins}(2016)}]{wiggins16}%
  \BibitemOpen
  \bibfield  {author} {\bibinfo {author} {\bibfnamefont {S.}~\bibnamefont
  {Wiggins}},\ }\href {\doibase 10.1134/S1560354716060034} {\bibfield
  {journal} {\bibinfo  {journal} {Regul. Chaotic Dyn.}\ }\textbf {\bibinfo
  {volume} {21}},\ \bibinfo {pages} {621} (\bibinfo {year} {2016}),\
  doi:10.1134/S1560354716060034}\BibitemShut {NoStop}%
\bibitem [{\citenamefont {Ezra}\ \emph {et~al.}(2009)\citenamefont {Ezra},
  \citenamefont {Waalkens},\ and\ \citenamefont {Wiggins}}]{Ezra2009}%
  \BibitemOpen
  \bibfield  {author} {\bibinfo {author} {\bibfnamefont {G.~S.}\ \bibnamefont
  {Ezra}}, \bibinfo {author} {\bibfnamefont {H.}~\bibnamefont {Waalkens}}, \
  and\ \bibinfo {author} {\bibfnamefont {S.}~\bibnamefont {Wiggins}},\ }\href
  {\doibase 10.1063/1.3119365} {\bibfield  {journal} {\bibinfo  {journal} {J.
  Chem. Phys.}\ }\textbf {\bibinfo {volume} {130}},\ \bibinfo {pages} {164118}
  (\bibinfo {year} {2009}),\ doi:10.1063/1.3119365}\BibitemShut {NoStop}%
\bibitem [{\citenamefont {Fenichel}(1972)}]{Fenichel72}%
  \BibitemOpen
  \bibfield  {author} {\bibinfo {author} {\bibfnamefont {N.}~\bibnamefont
  {Fenichel}},\ }\href {\doibase 10.1512/iumj.1972.21.21017} {\bibfield
  {journal} {\bibinfo  {journal} {Indiana Univ. Math. J.}\ }\textbf {\bibinfo
  {volume} {21}},\ \bibinfo {pages} {193} (\bibinfo {year} {1972}),\
  doi:10.1512/iumj.1972.21.21017}\BibitemShut {NoStop}%
\bibitem [{\citenamefont {Wiggins}(1994)}]{Wiggins94}%
  \BibitemOpen
  \bibfield  {author} {\bibinfo {author} {\bibfnamefont {S.}~\bibnamefont
  {Wiggins}},\ }\href@noop {} {\emph {\bibinfo {title} {Normally Hyperbolic
  Invariant Manifolds in Dynamical Systems}}}\ (\bibinfo  {publisher}
  {Springer},\ \bibinfo {address} {New York},\ \bibinfo {year}
  {1994})\BibitemShut {NoStop}%
\bibitem [{\citenamefont {Feldmaier}\ \emph {et~al.}(2017)\citenamefont
  {Feldmaier}, \citenamefont {Junginger}, \citenamefont {Main}, \citenamefont
  {Wunner},\ and\ \citenamefont {Hernandez}}]{hern17h}%
  \BibitemOpen
  \bibfield  {author} {\bibinfo {author} {\bibfnamefont {M.}~\bibnamefont
  {Feldmaier}}, \bibinfo {author} {\bibfnamefont {A.}~\bibnamefont
  {Junginger}}, \bibinfo {author} {\bibfnamefont {J.}~\bibnamefont {Main}},
  \bibinfo {author} {\bibfnamefont {G.}~\bibnamefont {Wunner}}, \ and\ \bibinfo
  {author} {\bibfnamefont {R.}~\bibnamefont {Hernandez}},\ }\href {\doibase
  10.1016/j.cplett.2017.09.008} {\bibfield  {journal} {\bibinfo  {journal}
  {Chem. Phys. Lett.}\ }\textbf {\bibinfo {volume} {687}},\ \bibinfo {pages}
  {194} (\bibinfo {year} {2017}),\
  doi:10.1016/j.cplett.2017.09.008}\BibitemShut {NoStop}%
\bibitem [{\citenamefont {Feldmaier}\ \emph
  {et~al.}(2019{\natexlab{a}})\citenamefont {Feldmaier}, \citenamefont
  {Schraft}, \citenamefont {Bardakcioglu}, \citenamefont {Reiff}, \citenamefont
  {Lober}, \citenamefont {Tsch{\"o}pe}, \citenamefont {Junginger},
  \citenamefont {Main}, \citenamefont {Bartsch},\ and\ \citenamefont
  {Hernandez}}]{hern19a}%
  \BibitemOpen
  \bibfield  {author} {\bibinfo {author} {\bibfnamefont {M.}~\bibnamefont
  {Feldmaier}}, \bibinfo {author} {\bibfnamefont {P.}~\bibnamefont {Schraft}},
  \bibinfo {author} {\bibfnamefont {R.}~\bibnamefont {Bardakcioglu}}, \bibinfo
  {author} {\bibfnamefont {J.}~\bibnamefont {Reiff}}, \bibinfo {author}
  {\bibfnamefont {M.}~\bibnamefont {Lober}}, \bibinfo {author} {\bibfnamefont
  {M.}~\bibnamefont {Tsch{\"o}pe}}, \bibinfo {author} {\bibfnamefont
  {A.}~\bibnamefont {Junginger}}, \bibinfo {author} {\bibfnamefont
  {J.}~\bibnamefont {Main}}, \bibinfo {author} {\bibfnamefont {T.}~\bibnamefont
  {Bartsch}}, \ and\ \bibinfo {author} {\bibfnamefont {R.}~\bibnamefont
  {Hernandez}},\ }\href {\doibase 10.1021/acs.jpcb.8b10541} {\bibfield
  {journal} {\bibinfo  {journal} {J. Phys. Chem. B}\ }\textbf {\bibinfo
  {volume} {123}},\ \bibinfo {pages} {2070} (\bibinfo {year}
  {2019}{\natexlab{a}}),\ doi:10.1021/acs.jpcb.8b10541}\BibitemShut {NoStop}%
\bibitem [{\citenamefont {Pollak}\ and\ \citenamefont
  {Pechukas}(1978)}]{pollak78}%
  \BibitemOpen
  \bibfield  {author} {\bibinfo {author} {\bibfnamefont {E.}~\bibnamefont
  {Pollak}}\ and\ \bibinfo {author} {\bibfnamefont {P.}~\bibnamefont
  {Pechukas}},\ }\href {\doibase 10.1063/1.436658} {\bibfield  {journal}
  {\bibinfo  {journal} {J. Chem. Phys.}\ }\textbf {\bibinfo {volume} {69}},\
  \bibinfo {pages} {1218} (\bibinfo {year} {1978}),\
  doi:10.1063/1.436658}\BibitemShut {NoStop}%
\bibitem [{\citenamefont {Pechukas}\ and\ \citenamefont
  {Pollak}(1979)}]{pech79a}%
  \BibitemOpen
  \bibfield  {author} {\bibinfo {author} {\bibfnamefont {P.}~\bibnamefont
  {Pechukas}}\ and\ \bibinfo {author} {\bibfnamefont {E.}~\bibnamefont
  {Pollak}},\ }\href {\doibase 10.1063/1.438575} {\bibfield  {journal}
  {\bibinfo  {journal} {J. Chem. Phys.}\ }\textbf {\bibinfo {volume} {71}},\
  \bibinfo {pages} {2062} (\bibinfo {year} {1979}),\
  doi:10.1063/1.438575}\BibitemShut {NoStop}%
\bibitem [{\citenamefont {Hernandez}\ and\ \citenamefont
  {Miller}(1993)}]{hern93b}%
  \BibitemOpen
  \bibfield  {author} {\bibinfo {author} {\bibfnamefont {R.}~\bibnamefont
  {Hernandez}}\ and\ \bibinfo {author} {\bibfnamefont {W.~H.}\ \bibnamefont
  {Miller}},\ }\href {\doibase 10.1016/0009-2614(93)90071-8} {\bibfield
  {journal} {\bibinfo  {journal} {Chem. Phys. Lett.}\ }\textbf {\bibinfo
  {volume} {214}},\ \bibinfo {pages} {129} (\bibinfo {year} {1993}),\
  doi:10.1016/0009-2614(93)90071-8}\BibitemShut {NoStop}%
\bibitem [{\citenamefont {Hernandez}(1994)}]{hern94}%
  \BibitemOpen
  \bibfield  {author} {\bibinfo {author} {\bibfnamefont {R.}~\bibnamefont
  {Hernandez}},\ }\href {\doibase 10.1063/1.467985} {\bibfield  {journal}
  {\bibinfo  {journal} {J. Chem. Phys.}\ }\textbf {\bibinfo {volume} {101}},\
  \bibinfo {pages} {9534} (\bibinfo {year} {1994}),\
  doi:10.1063/1.467985}\BibitemShut {NoStop}%
\bibitem [{\citenamefont {Wiggins}\ \emph {et~al.}(2001)\citenamefont
  {Wiggins}, \citenamefont {Wiesenfeld}, \citenamefont {Jaffe},\ and\
  \citenamefont {Uzer}}]{Wiggins01}%
  \BibitemOpen
  \bibfield  {author} {\bibinfo {author} {\bibfnamefont {S.}~\bibnamefont
  {Wiggins}}, \bibinfo {author} {\bibfnamefont {L.}~\bibnamefont {Wiesenfeld}},
  \bibinfo {author} {\bibfnamefont {C.}~\bibnamefont {Jaffe}}, \ and\ \bibinfo
  {author} {\bibfnamefont {T.}~\bibnamefont {Uzer}},\ }\href {\doibase
  10.1103/PhysRevLett.86.5478} {\bibfield  {journal} {\bibinfo  {journal}
  {Phys. Rev. Lett.}\ }\textbf {\bibinfo {volume} {86}} (\bibinfo {year}
  {2001}),\ doi:10.1103/PhysRevLett.86.5478}\BibitemShut {NoStop}%
\bibitem [{\citenamefont {Jaff{\'e}}\ \emph {et~al.}(2002)\citenamefont
  {Jaff{\'e}}, \citenamefont {Ross}, \citenamefont {Lo}, \citenamefont
  {Marsden}, \citenamefont {Farrelly},\ and\ \citenamefont {Uzer}}]{Jaffe02}%
  \BibitemOpen
  \bibfield  {author} {\bibinfo {author} {\bibfnamefont {C.}~\bibnamefont
  {Jaff{\'e}}}, \bibinfo {author} {\bibfnamefont {S.~D.}\ \bibnamefont {Ross}},
  \bibinfo {author} {\bibfnamefont {M.~W.}\ \bibnamefont {Lo}}, \bibinfo
  {author} {\bibfnamefont {J.}~\bibnamefont {Marsden}}, \bibinfo {author}
  {\bibfnamefont {D.}~\bibnamefont {Farrelly}}, \ and\ \bibinfo {author}
  {\bibfnamefont {T.}~\bibnamefont {Uzer}},\ }\href {\doibase
  10.1103/PhysRevLett.89.011101} {\bibfield  {journal} {\bibinfo  {journal}
  {Phys. Rev. Lett.}\ }\textbf {\bibinfo {volume} {89}},\ \bibinfo {pages}
  {011101} (\bibinfo {year} {2002}),\
  doi:10.1103/PhysRevLett.89.011101}\BibitemShut {NoStop}%
\bibitem [{\citenamefont {Teramoto}\ \emph {et~al.}(2011)\citenamefont
  {Teramoto}, \citenamefont {Toda},\ and\ \citenamefont
  {Komatsuzaki}}]{komatsuzaki11}%
  \BibitemOpen
  \bibfield  {author} {\bibinfo {author} {\bibfnamefont {H.}~\bibnamefont
  {Teramoto}}, \bibinfo {author} {\bibfnamefont {M.}~\bibnamefont {Toda}}, \
  and\ \bibinfo {author} {\bibfnamefont {T.}~\bibnamefont {Komatsuzaki}},\
  }\href {\doibase 10.1103/PhysRevLett.106.054101} {\bibfield  {journal}
  {\bibinfo  {journal} {Phys. Rev. Lett.}\ }\textbf {\bibinfo {volume} {106}},\
  \bibinfo {pages} {054101} (\bibinfo {year} {2011}),\
  doi:10.1103/PhysRevLett.106.054101}\BibitemShut {NoStop}%
\bibitem [{\citenamefont {Li}\ \emph {et~al.}(2006)\citenamefont {Li},
  \citenamefont {Shoujiguchi}, \citenamefont {Toda},\ and\ \citenamefont
  {Komatsuzaki}}]{komatsuzaki06a}%
  \BibitemOpen
  \bibfield  {author} {\bibinfo {author} {\bibfnamefont {C.-B.}\ \bibnamefont
  {Li}}, \bibinfo {author} {\bibfnamefont {A.}~\bibnamefont {Shoujiguchi}},
  \bibinfo {author} {\bibfnamefont {M.}~\bibnamefont {Toda}}, \ and\ \bibinfo
  {author} {\bibfnamefont {T.}~\bibnamefont {Komatsuzaki}},\ }\href {\doibase
  10.1103/PhysRevLett.97.028302} {\bibfield  {journal} {\bibinfo  {journal}
  {Phys. Rev. Lett.}\ }\textbf {\bibinfo {volume} {97}},\ \bibinfo {pages}
  {028302} (\bibinfo {year} {2006}),\
  doi:10.1103/PhysRevLett.97.028302}\BibitemShut {NoStop}%
\bibitem [{\citenamefont {Waalkens}\ and\ \citenamefont
  {Wiggins}(2004)}]{Waalkens04b}%
  \BibitemOpen
  \bibfield  {author} {\bibinfo {author} {\bibfnamefont {H.}~\bibnamefont
  {Waalkens}}\ and\ \bibinfo {author} {\bibfnamefont {S.}~\bibnamefont
  {Wiggins}},\ }\href {\doibase 10.1088/0305-4470/37/35/L02} {\bibfield
  {journal} {\bibinfo  {journal} {J. Phys. A}\ }\textbf {\bibinfo {volume}
  {37}},\ \bibinfo {pages} {L435} (\bibinfo {year} {2004}),\
  doi:10.1088/0305-4470/37/35/L02}\BibitemShut {NoStop}%
\bibitem [{\citenamefont {\ifmmode \mbox{\c{C}}\else
  \c{C}\fi{}ift\ifmmode~\mbox{\c{c}}\else \c{c}\fi{}i}\ and\ \citenamefont
  {Waalkens}(2013)}]{Waalkens13}%
  \BibitemOpen
  \bibfield  {author} {\bibinfo {author} {\bibfnamefont {U.}~\bibnamefont
  {\ifmmode \mbox{\c{C}}\else \c{C}\fi{}ift\ifmmode~\mbox{\c{c}}\else
  \c{c}\fi{}i}}\ and\ \bibinfo {author} {\bibfnamefont {H.}~\bibnamefont
  {Waalkens}},\ }\href {\doibase 10.1103/PhysRevLett.110.233201} {\bibfield
  {journal} {\bibinfo  {journal} {Phys. Rev. Lett.}\ }\textbf {\bibinfo
  {volume} {110}},\ \bibinfo {pages} {233201} (\bibinfo {year} {2013}),\
  doi:10.1103/PhysRevLett.110.233201}\BibitemShut {NoStop}%
\bibitem [{\citenamefont {Mancho}\ \emph {et~al.}(2003)\citenamefont {Mancho},
  \citenamefont {Small}, \citenamefont {Wiggins},\ and\ \citenamefont
  {Ide}}]{Mancho2003}%
  \BibitemOpen
  \bibfield  {author} {\bibinfo {author} {\bibfnamefont {A.~M.}\ \bibnamefont
  {Mancho}}, \bibinfo {author} {\bibfnamefont {D.}~\bibnamefont {Small}},
  \bibinfo {author} {\bibfnamefont {S.}~\bibnamefont {Wiggins}}, \ and\
  \bibinfo {author} {\bibfnamefont {K.}~\bibnamefont {Ide}},\ }\href {\doibase
  10.1016/S0167-2789(03)00152-0} {\bibfield  {journal} {\bibinfo  {journal}
  {Physica D}\ }\textbf {\bibinfo {volume} {182}},\ \bibinfo {pages} {188 }
  (\bibinfo {year} {2003}),\ doi:10.1016/S0167-2789(03)00152-0}\BibitemShut
  {NoStop}%
\bibitem [{\citenamefont {Craven}\ and\ \citenamefont
  {Hernandez}(2015)}]{hern15e}%
  \BibitemOpen
  \bibfield  {author} {\bibinfo {author} {\bibfnamefont {G.~T.}\ \bibnamefont
  {Craven}}\ and\ \bibinfo {author} {\bibfnamefont {R.}~\bibnamefont
  {Hernandez}},\ }\href {\doibase 10.1103/PhysRevLett.115.148301} {\bibfield
  {journal} {\bibinfo  {journal} {Phys. Rev. Lett.}\ }\textbf {\bibinfo
  {volume} {115}},\ \bibinfo {pages} {148301} (\bibinfo {year} {2015}),\
  doi:10.1103/PhysRevLett.115.148301}\BibitemShut {NoStop}%
\bibitem [{\citenamefont {Lopesino}\ \emph {et~al.}(2017)\citenamefont
  {Lopesino}, \citenamefont {Balibrea-Iniesta}, \citenamefont
  {Garc{\'{i}}a-Garrido}, \citenamefont {Wiggins},\ and\ \citenamefont
  {Mancho}}]{mancho17}%
  \BibitemOpen
  \bibfield  {author} {\bibinfo {author} {\bibfnamefont {C.}~\bibnamefont
  {Lopesino}}, \bibinfo {author} {\bibfnamefont {F.}~\bibnamefont
  {Balibrea-Iniesta}}, \bibinfo {author} {\bibfnamefont {V.~J.}\ \bibnamefont
  {Garc{\'{i}}a-Garrido}}, \bibinfo {author} {\bibfnamefont {S.}~\bibnamefont
  {Wiggins}}, \ and\ \bibinfo {author} {\bibfnamefont {A.~M.}\ \bibnamefont
  {Mancho}},\ }\href {\doibase 10.1142/S0218127417300014} {\bibfield  {journal}
  {\bibinfo  {journal} {Int. J. Bifurc. Chaos}\ }\textbf {\bibinfo {volume}
  {27}},\ \bibinfo {pages} {1730001} (\bibinfo {year} {2017}),\
  doi:10.1142/S0218127417300014}\BibitemShut {NoStop}%
\bibitem [{\citenamefont {Bardakcioglu}\ \emph {et~al.}(2018)\citenamefont
  {Bardakcioglu}, \citenamefont {Junginger}, \citenamefont {Feldmaier},
  \citenamefont {Main},\ and\ \citenamefont {Hernandez}}]{hern18g}%
  \BibitemOpen
  \bibfield  {author} {\bibinfo {author} {\bibfnamefont {R.}~\bibnamefont
  {Bardakcioglu}}, \bibinfo {author} {\bibfnamefont {A.}~\bibnamefont
  {Junginger}}, \bibinfo {author} {\bibfnamefont {M.}~\bibnamefont
  {Feldmaier}}, \bibinfo {author} {\bibfnamefont {J.}~\bibnamefont {Main}}, \
  and\ \bibinfo {author} {\bibfnamefont {R.}~\bibnamefont {Hernandez}},\ }\href
  {\doibase 10.1103/PhysRevE.98.032204} {\bibfield  {journal} {\bibinfo
  {journal} {Phys. Rev. E}\ }\textbf {\bibinfo {volume} {98}},\ \bibinfo
  {pages} {032204} (\bibinfo {year} {2018}),\
  doi:10.1103/PhysRevE.98.032204}\BibitemShut {NoStop}%
\bibitem [{\citenamefont {Schraft}\ \emph {et~al.}(2018)\citenamefont
  {Schraft}, \citenamefont {Junginger}, \citenamefont {Feldmaier},
  \citenamefont {Bardakcioglu}, \citenamefont {Main}, \citenamefont {Wunner},\
  and\ \citenamefont {Hernandez}}]{hern18c}%
  \BibitemOpen
  \bibfield  {author} {\bibinfo {author} {\bibfnamefont {P.}~\bibnamefont
  {Schraft}}, \bibinfo {author} {\bibfnamefont {A.}~\bibnamefont {Junginger}},
  \bibinfo {author} {\bibfnamefont {M.}~\bibnamefont {Feldmaier}}, \bibinfo
  {author} {\bibfnamefont {R.}~\bibnamefont {Bardakcioglu}}, \bibinfo {author}
  {\bibfnamefont {J.}~\bibnamefont {Main}}, \bibinfo {author} {\bibfnamefont
  {G.}~\bibnamefont {Wunner}}, \ and\ \bibinfo {author} {\bibfnamefont
  {R.}~\bibnamefont {Hernandez}},\ }\href {\doibase 10.1103/PhysRevE.97.042309}
  {\bibfield  {journal} {\bibinfo  {journal} {Phys. Rev. E}\ }\textbf {\bibinfo
  {volume} {97}},\ \bibinfo {pages} {042309} (\bibinfo {year} {2018}),\
  doi:10.1103/PhysRevE.97.042309}\BibitemShut {NoStop}%
\bibitem [{\citenamefont {Tsch{\"o}pe}\ \emph {et~al.}(2020)\citenamefont
  {Tsch{\"o}pe}, \citenamefont {Feldmaier}, \citenamefont {Main},\ and\
  \citenamefont {Hernandez}}]{hern20d}%
  \BibitemOpen
  \bibfield  {author} {\bibinfo {author} {\bibfnamefont {M.}~\bibnamefont
  {Tsch{\"o}pe}}, \bibinfo {author} {\bibfnamefont {M.}~\bibnamefont
  {Feldmaier}}, \bibinfo {author} {\bibfnamefont {J.}~\bibnamefont {Main}}, \
  and\ \bibinfo {author} {\bibfnamefont {R.}~\bibnamefont {Hernandez}},\ }\href
  {\doibase 10.1103/PhysRevE.101.022219} {\bibfield  {journal} {\bibinfo
  {journal} {Phys. Rev. E}\ }\textbf {\bibinfo {volume} {101}},\ \bibinfo
  {pages} {022219} (\bibinfo {year} {2020}),\
  doi:10.1103/PhysRevE.101.022219}\BibitemShut {NoStop}%
\bibitem [{\citenamefont {Revuelta}\ \emph {et~al.}(2017)\citenamefont
  {Revuelta}, \citenamefont {Craven}, \citenamefont {Bartsch}, \citenamefont
  {Borondo}, \citenamefont {Benito},\ and\ \citenamefont
  {Hernandez}}]{hern17f}%
  \BibitemOpen
  \bibfield  {author} {\bibinfo {author} {\bibfnamefont {F.}~\bibnamefont
  {Revuelta}}, \bibinfo {author} {\bibfnamefont {G.~T.}\ \bibnamefont
  {Craven}}, \bibinfo {author} {\bibfnamefont {T.}~\bibnamefont {Bartsch}},
  \bibinfo {author} {\bibfnamefont {F.}~\bibnamefont {Borondo}}, \bibinfo
  {author} {\bibfnamefont {R.~M.}\ \bibnamefont {Benito}}, \ and\ \bibinfo
  {author} {\bibfnamefont {R.}~\bibnamefont {Hernandez}},\ }\href {\doibase
  10.1063/1.4997571} {\bibfield  {journal} {\bibinfo  {journal} {J. Chem.
  Phys.}\ }\textbf {\bibinfo {volume} {147}},\ \bibinfo {pages} {074104}
  (\bibinfo {year} {2017}),\ doi:10.1063/1.4997571}\BibitemShut {NoStop}%
\bibitem [{\citenamefont {Bartsch}\ \emph {et~al.}(2019)\citenamefont
  {Bartsch}, \citenamefont {Revuelta}, \citenamefont {Benito},\ and\
  \citenamefont {Borondo}}]{bartsch19}%
  \BibitemOpen
  \bibfield  {author} {\bibinfo {author} {\bibfnamefont {T.}~\bibnamefont
  {Bartsch}}, \bibinfo {author} {\bibfnamefont {F.}~\bibnamefont {Revuelta}},
  \bibinfo {author} {\bibfnamefont {R.~M.}\ \bibnamefont {Benito}}, \ and\
  \bibinfo {author} {\bibfnamefont {F.}~\bibnamefont {Borondo}},\ }\href
  {\doibase 0.1103/PhysRevE.99.052211} {\bibfield  {journal} {\bibinfo
  {journal} {Phys. Rev. E}\ }\textbf {\bibinfo {volume} {99}},\ \bibinfo
  {pages} {052211} (\bibinfo {year} {2019}),\
  doi:0.1103/PhysRevE.99.052211}\BibitemShut {NoStop}%
\bibitem [{\citenamefont {Feldmaier}\ \emph
  {et~al.}(2019{\natexlab{b}})\citenamefont {Feldmaier}, \citenamefont
  {Bardakcioglu}, \citenamefont {Reiff}, \citenamefont {Main},\ and\
  \citenamefont {Hernandez}}]{hern19e}%
  \BibitemOpen
  \bibfield  {author} {\bibinfo {author} {\bibfnamefont {M.}~\bibnamefont
  {Feldmaier}}, \bibinfo {author} {\bibfnamefont {R.}~\bibnamefont
  {Bardakcioglu}}, \bibinfo {author} {\bibfnamefont {J.}~\bibnamefont {Reiff}},
  \bibinfo {author} {\bibfnamefont {J.}~\bibnamefont {Main}}, \ and\ \bibinfo
  {author} {\bibfnamefont {R.}~\bibnamefont {Hernandez}},\ }\href {\doibase
  10.1063/1.5127539} {\bibfield  {journal} {\bibinfo  {journal} {J. Chem.
  Phys.}\ }\textbf {\bibinfo {volume} {151}},\ \bibinfo {pages} {244108}
  (\bibinfo {year} {2019}{\natexlab{b}}),\ doi:10.1063/1.5127539}\BibitemShut
  {NoStop}%
\bibitem [{\citenamefont {Craven}\ \emph {et~al.}(2014)\citenamefont {Craven},
  \citenamefont {Bartsch},\ and\ \citenamefont {Hernandez}}]{hern14f}%
  \BibitemOpen
  \bibfield  {author} {\bibinfo {author} {\bibfnamefont {G.~T.}\ \bibnamefont
  {Craven}}, \bibinfo {author} {\bibfnamefont {T.}~\bibnamefont {Bartsch}}, \
  and\ \bibinfo {author} {\bibfnamefont {R.}~\bibnamefont {Hernandez}},\ }\href
  {\doibase 10.1063/1.4891471} {\bibfield  {journal} {\bibinfo  {journal} {J.
  Chem. Phys.}\ }\textbf {\bibinfo {volume} {141}},\ \bibinfo {pages} {041106}
  (\bibinfo {year} {2014}),\ doi:10.1063/1.4891471}\BibitemShut {NoStop}%
\bibitem [{\citenamefont {Bartsch}\ \emph
  {et~al.}(2005{\natexlab{b}})\citenamefont {Bartsch}, \citenamefont {Uzer},\
  and\ \citenamefont {Hernandez}}]{dawn05b}%
  \BibitemOpen
  \bibfield  {author} {\bibinfo {author} {\bibfnamefont {T.}~\bibnamefont
  {Bartsch}}, \bibinfo {author} {\bibfnamefont {T.}~\bibnamefont {Uzer}}, \
  and\ \bibinfo {author} {\bibfnamefont {R.}~\bibnamefont {Hernandez}},\ }\href
  {\doibase 10.1063/1.2109827} {\bibfield  {journal} {\bibinfo  {journal} {J.
  Chem. Phys.}\ }\textbf {\bibinfo {volume} {123}},\ \bibinfo {pages} {204102}
  (\bibinfo {year} {2005}{\natexlab{b}}),\ doi:10.1063/1.2109827}\BibitemShut
  {NoStop}%
\bibitem [{\citenamefont {Truhlar}\ \emph {et~al.}(1982)\citenamefont
  {Truhlar}, \citenamefont {Isaacson}, \citenamefont {Skodje},\ and\
  \citenamefont {Garrett}}]{truh82}%
  \BibitemOpen
  \bibfield  {author} {\bibinfo {author} {\bibfnamefont {D.~G.}\ \bibnamefont
  {Truhlar}}, \bibinfo {author} {\bibfnamefont {A.~D.}\ \bibnamefont
  {Isaacson}}, \bibinfo {author} {\bibfnamefont {R.~T.}\ \bibnamefont
  {Skodje}}, \ and\ \bibinfo {author} {\bibfnamefont {B.~C.}\ \bibnamefont
  {Garrett}},\ }\href {\doibase 10.1021/j100209a021} {\bibfield  {journal}
  {\bibinfo  {journal} {J. Phys. Chem.}\ }\textbf {\bibinfo {volume} {86}},\
  \bibinfo {pages} {2252} (\bibinfo {year} {1982}),\
  doi:10.1021/j100209a021}\BibitemShut {NoStop}%
\bibitem [{\citenamefont {Pollak}(2000)}]{Pollak00}%
  \BibitemOpen
  \bibfield  {author} {\bibinfo {author} {\bibfnamefont {E.}~\bibnamefont
  {Pollak}},\ }in\ \href@noop {} {\emph {\bibinfo {booktitle} {Theoretical
  Methods in Condensed Phase Chemistry}}},\ \bibinfo {editor} {edited by\
  \bibinfo {editor} {\bibfnamefont {S.~D.}\ \bibnamefont {Schwartz}}}\
  (\bibinfo  {publisher} {Kluwer Academic Publishers},\ \bibinfo {address}
  {Dordrecht},\ \bibinfo {year} {2000})\ pp.\ \bibinfo {pages}
  {1--46}\BibitemShut {NoStop}%
\bibitem [{\citenamefont {Baraban}\ \emph {et~al.}(2015)\citenamefont
  {Baraban}, \citenamefont {Changala}, \citenamefont {Mellau}, \citenamefont
  {Stanton}, \citenamefont {Merer},\ and\ \citenamefont {Field}}]{stanton17}%
  \BibitemOpen
  \bibfield  {author} {\bibinfo {author} {\bibfnamefont {J.~H.}\ \bibnamefont
  {Baraban}}, \bibinfo {author} {\bibfnamefont {P.~B.}\ \bibnamefont
  {Changala}}, \bibinfo {author} {\bibfnamefont {G.~C.}\ \bibnamefont
  {Mellau}}, \bibinfo {author} {\bibfnamefont {J.~F.}\ \bibnamefont {Stanton}},
  \bibinfo {author} {\bibfnamefont {A.~J.}\ \bibnamefont {Merer}}, \ and\
  \bibinfo {author} {\bibfnamefont {R.~W.}\ \bibnamefont {Field}},\ }\href
  {\doibase 10.1126/science.aac9668} {\bibfield  {journal} {\bibinfo  {journal}
  {Science}\ }\textbf {\bibinfo {volume} {350}},\ \bibinfo {pages} {1338}
  (\bibinfo {year} {2015}),\ doi:10.1126/science.aac9668}\BibitemShut {NoStop}%
\bibitem [{\citenamefont {Waalkens}\ \emph {et~al.}(2004)\citenamefont
  {Waalkens}, \citenamefont {Burbanks},\ and\ \citenamefont
  {Wiggins}}]{Waalkens04c}%
  \BibitemOpen
  \bibfield  {author} {\bibinfo {author} {\bibfnamefont {H.}~\bibnamefont
  {Waalkens}}, \bibinfo {author} {\bibfnamefont {A.}~\bibnamefont {Burbanks}},
  \ and\ \bibinfo {author} {\bibfnamefont {S.}~\bibnamefont {Wiggins}},\
  }\href@noop {} {\bibfield  {journal} {\bibinfo  {journal} {J. Chem. Phys.}\
  }\textbf {\bibinfo {volume} {121}},\ \bibinfo {pages} {6207} (\bibinfo {year}
  {2004})}\BibitemShut {NoStop}%
\bibitem [{\citenamefont {McCoy}\ and\ \citenamefont {{Sibert
  III}}(1991)}]{sibe91}%
  \BibitemOpen
  \bibfield  {author} {\bibinfo {author} {\bibfnamefont {A.~B.}\ \bibnamefont
  {McCoy}}\ and\ \bibinfo {author} {\bibfnamefont {E.~L.}\ \bibnamefont
  {{Sibert III}}},\ }\href {\doibase 10.1063/1.460851} {\bibfield  {journal}
  {\bibinfo  {journal} {J. Chem. Phys.}\ }\textbf {\bibinfo {volume} {95}},\
  \bibinfo {pages} {3476} (\bibinfo {year} {1991}),\
  doi:10.1063/1.460851}\BibitemShut {NoStop}%
\bibitem [{\citenamefont {Essers}\ \emph {et~al.}(1982)\citenamefont {Essers},
  \citenamefont {Tennyson},\ and\ \citenamefont {Wormer}}]{essers1982scf}%
  \BibitemOpen
  \bibfield  {author} {\bibinfo {author} {\bibfnamefont {R.}~\bibnamefont
  {Essers}}, \bibinfo {author} {\bibfnamefont {J.}~\bibnamefont {Tennyson}}, \
  and\ \bibinfo {author} {\bibfnamefont {P.~E.~S.}\ \bibnamefont {Wormer}},\
  }\href {\doibase 10.1016/0009-2614(82)80046-8} {\bibfield  {journal}
  {\bibinfo  {journal} {Chem. Phys. Lett.}\ }\textbf {\bibinfo {volume} {89}},\
  \bibinfo {pages} {223} (\bibinfo {year} {1982}),\
  doi:10.1016/0009-2614(82)80046-8}\BibitemShut {NoStop}%
\bibitem [{\citenamefont {Brocks}\ \emph {et~al.}(1984)\citenamefont {Brocks},
  \citenamefont {Tennyson},\ and\ \citenamefont {{van der
  Avoird}}}]{brocks1984abinitio}%
  \BibitemOpen
  \bibfield  {author} {\bibinfo {author} {\bibfnamefont {G.}~\bibnamefont
  {Brocks}}, \bibinfo {author} {\bibfnamefont {J.}~\bibnamefont {Tennyson}}, \
  and\ \bibinfo {author} {\bibfnamefont {A.}~\bibnamefont {{van der Avoird}}},\
  }\href {\doibase 10.1063/1.447075} {\bibfield  {journal} {\bibinfo  {journal}
  {J. Chem. Phys.}\ }\textbf {\bibinfo {volume} {80}},\ \bibinfo {pages} {3223}
  (\bibinfo {year} {1984}),\ doi:10.1063/1.447075}\BibitemShut {NoStop}%
\bibitem [{\citenamefont {{Garc{\'\i}a-M\"uller}}\ \emph
  {et~al.}(2008)\citenamefont {{Garc{\'\i}a-M\"uller}}, \citenamefont
  {Borondo}, \citenamefont {Hernandez},\ and\ \citenamefont
  {Benito}}]{hern08g}%
  \BibitemOpen
  \bibfield  {author} {\bibinfo {author} {\bibfnamefont {P.~L.}\ \bibnamefont
  {{Garc{\'\i}a-M\"uller}}}, \bibinfo {author} {\bibfnamefont {F.}~\bibnamefont
  {Borondo}}, \bibinfo {author} {\bibfnamefont {R.}~\bibnamefont {Hernandez}},
  \ and\ \bibinfo {author} {\bibfnamefont {R.~M.}\ \bibnamefont {Benito}},\
  }\href {\doibase 10.1103/PhysRevLett.101.178302} {\bibfield  {journal}
  {\bibinfo  {journal} {Phys. Rev. Lett.}\ }\textbf {\bibinfo {volume} {101}},\
  \bibinfo {pages} {178302} (\bibinfo {year} {2008}),\
  doi:10.1103/PhysRevLett.101.178302}\BibitemShut {NoStop}%
\bibitem [{\citenamefont {{Garc{\'\i}a-M\"uller}}\ \emph
  {et~al.}(2012)\citenamefont {{Garc{\'\i}a-M\"uller}}, \citenamefont
  {Hernandez}, \citenamefont {Benito},\ and\ \citenamefont
  {Borondo}}]{hern12e}%
  \BibitemOpen
  \bibfield  {author} {\bibinfo {author} {\bibfnamefont {P.~L.}\ \bibnamefont
  {{Garc{\'\i}a-M\"uller}}}, \bibinfo {author} {\bibfnamefont {R.}~\bibnamefont
  {Hernandez}}, \bibinfo {author} {\bibfnamefont {R.~M.}\ \bibnamefont
  {Benito}}, \ and\ \bibinfo {author} {\bibfnamefont {F.}~\bibnamefont
  {Borondo}},\ }\href {\doibase 10.1063/1.4766257} {\bibfield  {journal}
  {\bibinfo  {journal} {J. Chem. Phys.}\ }\textbf {\bibinfo {volume} {137}},\
  \bibinfo {pages} {204301} (\bibinfo {year} {2012}),\
  doi:10.1063/1.4766257}\BibitemShut {NoStop}%
\bibitem [{\citenamefont {{Garc{\'\i}a-M\"uller}}\ \emph
  {et~al.}(2014)\citenamefont {{Garc{\'\i}a-M\"uller}}, \citenamefont
  {Hernandez}, \citenamefont {Benito},\ and\ \citenamefont
  {Borondo}}]{hern14j}%
  \BibitemOpen
  \bibfield  {author} {\bibinfo {author} {\bibfnamefont {P.~L.}\ \bibnamefont
  {{Garc{\'\i}a-M\"uller}}}, \bibinfo {author} {\bibfnamefont {R.}~\bibnamefont
  {Hernandez}}, \bibinfo {author} {\bibfnamefont {R.~M.}\ \bibnamefont
  {Benito}}, \ and\ \bibinfo {author} {\bibfnamefont {F.}~\bibnamefont
  {Borondo}},\ }\href {\doibase 10.1063/1.4892921} {\bibfield  {journal}
  {\bibinfo  {journal} {J. Chem. Phys.}\ }\textbf {\bibinfo {volume} {141}},\
  \bibinfo {pages} {074312} (\bibinfo {year} {2014}),\
  doi:10.1063/1.4892921}\BibitemShut {NoStop}%
\bibitem [{\citenamefont {Junginger}\ \emph {et~al.}(2016)\citenamefont
  {Junginger}, \citenamefont {{Garc{\'\i}a-M\"uller}}, \citenamefont {Borondo},
  \citenamefont {Benito},\ and\ \citenamefont {Hernandez}}]{hern16c}%
  \BibitemOpen
  \bibfield  {author} {\bibinfo {author} {\bibfnamefont {A.}~\bibnamefont
  {Junginger}}, \bibinfo {author} {\bibfnamefont {P.~L.}\ \bibnamefont
  {{Garc{\'\i}a-M\"uller}}}, \bibinfo {author} {\bibfnamefont {F.}~\bibnamefont
  {Borondo}}, \bibinfo {author} {\bibfnamefont {R.~M.}\ \bibnamefont {Benito}},
  \ and\ \bibinfo {author} {\bibfnamefont {R.}~\bibnamefont {Hernandez}},\
  }\href {\doibase 10.1063/1.4939480} {\bibfield  {journal} {\bibinfo
  {journal} {J. Chem. Phys.}\ }\textbf {\bibinfo {volume} {144}},\ \bibinfo
  {pages} {024104} (\bibinfo {year} {2016}),\
  doi:10.1063/1.4939480}\BibitemShut {NoStop}%
\bibitem [{\citenamefont {Vergel}\ \emph {et~al.}(2014)\citenamefont {Vergel},
  \citenamefont {Benito}, \citenamefont {Losada},\ and\ \citenamefont
  {Borondo}}]{vergel2014geometrical}%
  \BibitemOpen
  \bibfield  {author} {\bibinfo {author} {\bibfnamefont {A.}~\bibnamefont
  {Vergel}}, \bibinfo {author} {\bibfnamefont {R.~M.}\ \bibnamefont {Benito}},
  \bibinfo {author} {\bibfnamefont {J.~C.}\ \bibnamefont {Losada}}, \ and\
  \bibinfo {author} {\bibfnamefont {F.}~\bibnamefont {Borondo}},\ }\href
  {\doibase 10.1103/PhysRevE.89.022901} {\bibfield  {journal} {\bibinfo
  {journal} {Phys. Rev. E}\ }\textbf {\bibinfo {volume} {89}},\ \bibinfo
  {pages} {022901} (\bibinfo {year} {2014}),\
  doi:10.1103/PhysRevE.89.022901}\BibitemShut {NoStop}%
\bibitem [{\citenamefont {Prado}\ \emph {et~al.}(2009)\citenamefont {Prado},
  \citenamefont {Vergini}, \citenamefont {Benito},\ and\ \citenamefont
  {Borondo}}]{Prado2009}%
  \BibitemOpen
  \bibfield  {author} {\bibinfo {author} {\bibfnamefont {S.~D.}\ \bibnamefont
  {Prado}}, \bibinfo {author} {\bibfnamefont {E.~G.}\ \bibnamefont {Vergini}},
  \bibinfo {author} {\bibfnamefont {R.~M.}\ \bibnamefont {Benito}}, \ and\
  \bibinfo {author} {\bibfnamefont {F.}~\bibnamefont {Borondo}},\ }\href
  {\doibase 10.1209/0295-5075/88/40003} {\bibfield  {journal} {\bibinfo
  {journal} {Europhys. Lett.}\ }\textbf {\bibinfo {volume} {88}},\ \bibinfo
  {pages} {40003} (\bibinfo {year} {2009}),\
  doi:10.1209/0295-5075/88/40003}\BibitemShut {NoStop}%
\bibitem [{\citenamefont {Murgida}\ \emph {et~al.}(2010)\citenamefont
  {Murgida}, \citenamefont {Wisniacki}, \citenamefont {Tamborenea},\ and\
  \citenamefont {Borondo}}]{borondo10}%
  \BibitemOpen
  \bibfield  {author} {\bibinfo {author} {\bibfnamefont {G.~E.}\ \bibnamefont
  {Murgida}}, \bibinfo {author} {\bibfnamefont {D.~A.}\ \bibnamefont
  {Wisniacki}}, \bibinfo {author} {\bibfnamefont {P.~I.}\ \bibnamefont
  {Tamborenea}}, \ and\ \bibinfo {author} {\bibfnamefont {F.}~\bibnamefont
  {Borondo}},\ }\href {\doibase 10.1016/j.cplett.2010.07.057} {\bibfield
  {journal} {\bibinfo  {journal} {Chem. Phys. Lett.}\ }\textbf {\bibinfo
  {volume} {496}},\ \bibinfo {pages} {356} (\bibinfo {year} {2010}),\
  doi:10.1016/j.cplett.2010.07.057}\BibitemShut {NoStop}%
\bibitem [{\citenamefont {Revuelta}\ \emph {et~al.}(2015)\citenamefont
  {Revuelta}, \citenamefont {Chac\'on},\ and\ \citenamefont
  {Borondo}}]{Revuelta2015}%
  \BibitemOpen
  \bibfield  {author} {\bibinfo {author} {\bibfnamefont {F.}~\bibnamefont
  {Revuelta}}, \bibinfo {author} {\bibfnamefont {R.}~\bibnamefont {Chac\'on}},
  \ and\ \bibinfo {author} {\bibfnamefont {F.}~\bibnamefont {Borondo}},\ }\href
  {\doibase 10.1209/0295-5075/110/40007} {\bibfield  {journal} {\bibinfo
  {journal} {Europhys. Lett.}\ }\textbf {\bibinfo {volume} {110}},\ \bibinfo
  {pages} {40007} (\bibinfo {year} {2015}),\
  doi:10.1209/0295-5075/110/40007}\BibitemShut {NoStop}%
\bibitem [{\citenamefont {Murgida}\ \emph {et~al.}(2015)\citenamefont
  {Murgida}, \citenamefont {Arranz},\ and\ \citenamefont
  {Borondo}}]{murgida2015quantum}%
  \BibitemOpen
  \bibfield  {author} {\bibinfo {author} {\bibfnamefont {G.~E.}\ \bibnamefont
  {Murgida}}, \bibinfo {author} {\bibfnamefont {F.~J.}\ \bibnamefont {Arranz}},
  \ and\ \bibinfo {author} {\bibfnamefont {F.}~\bibnamefont {Borondo}},\ }\href
  {\doibase 10.1063/1.4936424} {\bibfield  {journal} {\bibinfo  {journal} {J.
  Chem. Phys.}\ }\textbf {\bibinfo {volume} {143}},\ \bibinfo {pages} {214305}
  (\bibinfo {year} {2015}),\ doi:10.1063/1.4936424}\BibitemShut {NoStop}%
\bibitem [{\citenamefont {Brocks}\ and\ \citenamefont
  {Tennyson}(1983)}]{brocks1983ab}%
  \BibitemOpen
  \bibfield  {author} {\bibinfo {author} {\bibfnamefont {G.}~\bibnamefont
  {Brocks}}\ and\ \bibinfo {author} {\bibfnamefont {J.}~\bibnamefont
  {Tennyson}},\ }\href@noop {} {\bibfield  {journal} {\bibinfo  {journal} {J.
  Mol. Spectrosc.}\ }\textbf {\bibinfo {volume} {99}},\ \bibinfo {pages} {263}
  (\bibinfo {year} {1983})}\BibitemShut {NoStop}%
\bibitem [{\citenamefont {Wormer}\ and\ \citenamefont
  {Tennyson}(1981)}]{wormer1981abinitio}%
  \BibitemOpen
  \bibfield  {author} {\bibinfo {author} {\bibfnamefont {P.~E.~S.}\
  \bibnamefont {Wormer}}\ and\ \bibinfo {author} {\bibfnamefont
  {J.}~\bibnamefont {Tennyson}},\ }\href {\doibase 10.1063/1.442174} {\bibfield
   {journal} {\bibinfo  {journal} {J. Chem. Phys.}\ }\textbf {\bibinfo {volume}
  {75}},\ \bibinfo {pages} {1245} (\bibinfo {year} {1981}),\
  doi:10.1063/1.442174}\BibitemShut {NoStop}%
\bibitem [{\citenamefont {Benito}\ \emph {et~al.}(1989)\citenamefont {Benito},
  \citenamefont {Borondo}, \citenamefont {Kim}, \citenamefont {Sumpter},\ and\
  \citenamefont {Ezra}}]{borondo89a}%
  \BibitemOpen
  \bibfield  {author} {\bibinfo {author} {\bibfnamefont {R.}~\bibnamefont
  {Benito}}, \bibinfo {author} {\bibfnamefont {F.}~\bibnamefont {Borondo}},
  \bibinfo {author} {\bibfnamefont {J.-H.}\ \bibnamefont {Kim}}, \bibinfo
  {author} {\bibfnamefont {B.}~\bibnamefont {Sumpter}}, \ and\ \bibinfo
  {author} {\bibfnamefont {G.}~\bibnamefont {Ezra}},\ }\href {\doibase
  10.1016/S0009-2614(89)87032-0} {\bibfield  {journal} {\bibinfo  {journal}
  {Chem. Phys. Lett.}\ }\textbf {\bibinfo {volume} {161}},\ \bibinfo {pages}
  {60} (\bibinfo {year} {1989}),\
  doi:10.1016/S0009-2614(89)87032-0}\BibitemShut {NoStop}%
\bibitem [{\citenamefont {Press}\ \emph {et~al.}(1987)\citenamefont {Press},
  \citenamefont {Teukolsky}, \citenamefont {Vetterling},\ and\ \citenamefont
  {Flannery}}]{Press1987}%
  \BibitemOpen
  \bibfield  {author} {\bibinfo {author} {\bibfnamefont {W.~H.}\ \bibnamefont
  {Press}}, \bibinfo {author} {\bibfnamefont {S.~A.}\ \bibnamefont
  {Teukolsky}}, \bibinfo {author} {\bibfnamefont {W.~T.}\ \bibnamefont
  {Vetterling}}, \ and\ \bibinfo {author} {\bibfnamefont {B.~P.}\ \bibnamefont
  {Flannery}},\ }\href@noop {} {\emph {\bibinfo {title} {Numerical recipes: The
  art of scientific computing}}}\ (\bibinfo  {publisher} {Cambridge University
  Press},\ \bibinfo {address} {New York},\ \bibinfo {year} {1987})\BibitemShut
  {NoStop}%
\bibitem [{Note1()}]{Note1}%
  \BibitemOpen
  \bibinfo {note} {Throughout the paper, we use atomic units---Bohr radius \si
  {\bohr }, elementary charge \si {\elementarycharge }, and \SI {1}{\hartree }
  (Hartree)---for length, charge, and energy, but \SI {1}{\atomicmassunit }
  (dalton) as the mass unit. As a consequence, times must be multiplied and
  frequencies and rates must be divided by a factor of $\protect \sqrt {\SI
  {1}{\atomicmassunit } / \si {\electronmass }} = \protect \sqrt {1823} = 42.7$
  to obtain the corresponding values in atomic units.}\BibitemShut {Stop}%
\bibitem [{\citenamefont {Jackson}(2012)}]{jackson2012classical}%
  \BibitemOpen
  \bibfield  {author} {\bibinfo {author} {\bibfnamefont {J.~D.}\ \bibnamefont
  {Jackson}},\ }\href@noop {} {\emph {\bibinfo {title} {Classical
  electrodynamics}}}\ (\bibinfo  {publisher} {John Wiley \& Sons},\ \bibinfo
  {year} {2012})\BibitemShut {NoStop}%
\bibitem [{Note2()}]{Note2}%
  \BibitemOpen
  \bibinfo {note} {These values have been updated to reflect a correction due
  to the unit conversion noted in footnote~\protect \citenum
  {Note1}.}\BibitemShut {Stop}%
\bibitem [{\citenamefont {Wimberger}(2014)}]{wimberger2014nonlinear}%
  \BibitemOpen
  \bibfield  {author} {\bibinfo {author} {\bibfnamefont {S.}~\bibnamefont
  {Wimberger}},\ }\href@noop {} {\emph {\bibinfo {title} {Nonlinear dynamics
  and quantum chaos}}}\ (\bibinfo  {publisher} {Springer},\ \bibinfo {address}
  {Cham},\ \bibinfo {year} {2014})\BibitemShut {NoStop}%
\bibitem [{\citenamefont {Revuelta}\ \emph {et~al.}(2019)\citenamefont
  {Revuelta}, \citenamefont {Benito},\ and\ \citenamefont
  {Borondo}}]{Revuelta19}%
  \BibitemOpen
  \bibfield  {author} {\bibinfo {author} {\bibfnamefont {F.}~\bibnamefont
  {Revuelta}}, \bibinfo {author} {\bibfnamefont {R.~M.}\ \bibnamefont
  {Benito}}, \ and\ \bibinfo {author} {\bibfnamefont {F.}~\bibnamefont
  {Borondo}},\ }\href {\doibase 10.1103/PhysRevE.99.032221} {\bibfield
  {journal} {\bibinfo  {journal} {Phys. Rev. E}\ }\textbf {\bibinfo {volume}
  {99}},\ \bibinfo {pages} {032221} (\bibinfo {year} {2019}),\
  doi:10.1103/PhysRevE.99.032221}\BibitemShut {NoStop}%
\end{thebibliography}%


\begin{thebibliography}{9}%
\makeatletter
\providecommand \@ifxundefined [1]{%
 \@ifx{#1\undefined}
}%
\providecommand \@ifnum [1]{%
 \ifnum #1\expandafter \@firstoftwo
 \else \expandafter \@secondoftwo
 \fi
}%
\providecommand \@ifx [1]{%
 \ifx #1\expandafter \@firstoftwo
 \else \expandafter \@secondoftwo
 \fi
}%
\providecommand \natexlab [1]{#1}%
\providecommand \enquote  [1]{``#1''}%
\providecommand \bibnamefont  [1]{#1}%
\providecommand \bibfnamefont [1]{#1}%
\providecommand \citenamefont [1]{#1}%
\providecommand \href@noop [0]{\@secondoftwo}%
\providecommand \href [0]{\begingroup \@sanitize@url \@href}%
\providecommand \@href[1]{\@@startlink{#1}\@@href}%
\providecommand \@@href[1]{\endgroup#1\@@endlink}%
\providecommand \@sanitize@url [0]{\catcode `\\12\catcode `\$12\catcode
  `\&12\catcode `\#12\catcode `\^12\catcode `\_12\catcode `\%12\relax}%
\providecommand \@@startlink[1]{}%
\providecommand \@@endlink[0]{}%
\providecommand \url  [0]{\begingroup\@sanitize@url \@url }%
\providecommand \@url [1]{\endgroup\@href {#1}{\urlprefix }}%
\providecommand \urlprefix  [0]{URL }%
\providecommand \Eprint [0]{\href }%
\providecommand \doibase [0]{http://dx.doi.org/}%
\providecommand \selectlanguage [0]{\@gobble}%
\providecommand \bibinfo  [0]{\@secondoftwo}%
\providecommand \bibfield  [0]{\@secondoftwo}%
\providecommand \translation [1]{[#1]}%
\providecommand \BibitemOpen [0]{}%
\providecommand \bibitemStop [0]{}%
\providecommand \bibitemNoStop [0]{.\EOS\space}%
\providecommand \EOS [0]{\spacefactor3000\relax}%
\providecommand \BibitemShut  [1]{\csname bibitem#1\endcsname}%
\let\auto@bib@innerbib\@empty
\bibitem [{\citenamefont {Junginger}\ \emph {et~al.}(2017)\citenamefont
  {Junginger}, \citenamefont {Duvenbeck}, \citenamefont {Feldmaier},
  \citenamefont {Main}, \citenamefont {Wunner},\ and\ \citenamefont
  {Hernandez}}]{hern17e}%
  \BibitemOpen
  \bibfield  {author} {\bibinfo {author} {\bibfnamefont {A.}~\bibnamefont
  {Junginger}}, \bibinfo {author} {\bibfnamefont {L.}~\bibnamefont
  {Duvenbeck}}, \bibinfo {author} {\bibfnamefont {M.}~\bibnamefont
  {Feldmaier}}, \bibinfo {author} {\bibfnamefont {J.}~\bibnamefont {Main}},
  \bibinfo {author} {\bibfnamefont {G.}~\bibnamefont {Wunner}}, \ and\ \bibinfo
  {author} {\bibfnamefont {R.}~\bibnamefont {Hernandez}},\ }\href {\doibase
  10.1063/1.4997379} {\bibfield  {journal} {\bibinfo  {journal} {J. Chem.
  Phys.}\ }\textbf {\bibinfo {volume} {147}},\ \bibinfo {pages} {064101}
  (\bibinfo {year} {2017})}\BibitemShut {NoStop}%
\bibitem [{\citenamefont {Feldmaier}\ \emph
  {et~al.}(2019{\natexlab{a}})\citenamefont {Feldmaier}, \citenamefont
  {Schraft}, \citenamefont {Bardakcioglu}, \citenamefont {Reiff}, \citenamefont
  {Lober}, \citenamefont {Tsch{\"o}pe}, \citenamefont {Junginger},
  \citenamefont {Main}, \citenamefont {Bartsch},\ and\ \citenamefont
  {Hernandez}}]{hern19a}%
  \BibitemOpen
  \bibfield  {author} {\bibinfo {author} {\bibfnamefont {M.}~\bibnamefont
  {Feldmaier}}, \bibinfo {author} {\bibfnamefont {P.}~\bibnamefont {Schraft}},
  \bibinfo {author} {\bibfnamefont {R.}~\bibnamefont {Bardakcioglu}}, \bibinfo
  {author} {\bibfnamefont {J.}~\bibnamefont {Reiff}}, \bibinfo {author}
  {\bibfnamefont {M.}~\bibnamefont {Lober}}, \bibinfo {author} {\bibfnamefont
  {M.}~\bibnamefont {Tsch{\"o}pe}}, \bibinfo {author} {\bibfnamefont
  {A.}~\bibnamefont {Junginger}}, \bibinfo {author} {\bibfnamefont
  {J.}~\bibnamefont {Main}}, \bibinfo {author} {\bibfnamefont {T.}~\bibnamefont
  {Bartsch}}, \ and\ \bibinfo {author} {\bibfnamefont {R.}~\bibnamefont
  {Hernandez}},\ }\href {\doibase 10.1021/acs.jpcb.8b10541} {\bibfield
  {journal} {\bibinfo  {journal} {J. Phys. Chem. B}\ }\textbf {\bibinfo
  {volume} {123}},\ \bibinfo {pages} {2070} (\bibinfo {year}
  {2019}{\natexlab{a}})}\BibitemShut {NoStop}%
\bibitem [{\citenamefont {Bardakcioglu}\ \emph {et~al.}(2018)\citenamefont
  {Bardakcioglu}, \citenamefont {Junginger}, \citenamefont {Feldmaier},
  \citenamefont {Main},\ and\ \citenamefont {Hernandez}}]{hern18g}%
  \BibitemOpen
  \bibfield  {author} {\bibinfo {author} {\bibfnamefont {R.}~\bibnamefont
  {Bardakcioglu}}, \bibinfo {author} {\bibfnamefont {A.}~\bibnamefont
  {Junginger}}, \bibinfo {author} {\bibfnamefont {M.}~\bibnamefont
  {Feldmaier}}, \bibinfo {author} {\bibfnamefont {J.}~\bibnamefont {Main}}, \
  and\ \bibinfo {author} {\bibfnamefont {R.}~\bibnamefont {Hernandez}},\ }\href
  {\doibase 10.1103/PhysRevE.98.032204} {\bibfield  {journal} {\bibinfo
  {journal} {Phys. Rev. E}\ }\textbf {\bibinfo {volume} {98}},\ \bibinfo
  {pages} {032204} (\bibinfo {year} {2018})}\BibitemShut {NoStop}%
\bibitem [{\citenamefont {Feldmaier}\ \emph
  {et~al.}(2019{\natexlab{b}})\citenamefont {Feldmaier}, \citenamefont
  {Bardakcioglu}, \citenamefont {Reiff}, \citenamefont {Main},\ and\
  \citenamefont {Hernandez}}]{hern19e}%
  \BibitemOpen
  \bibfield  {author} {\bibinfo {author} {\bibfnamefont {M.}~\bibnamefont
  {Feldmaier}}, \bibinfo {author} {\bibfnamefont {R.}~\bibnamefont
  {Bardakcioglu}}, \bibinfo {author} {\bibfnamefont {J.}~\bibnamefont {Reiff}},
  \bibinfo {author} {\bibfnamefont {J.}~\bibnamefont {Main}}, \ and\ \bibinfo
  {author} {\bibfnamefont {R.}~\bibnamefont {Hernandez}},\ }\href {\doibase
  10.1063/1.5127539} {\bibfield  {journal} {\bibinfo  {journal} {J. Chem.
  Phys.}\ }\textbf {\bibinfo {volume} {151}},\ \bibinfo {pages} {244108}
  (\bibinfo {year} {2019}{\natexlab{b}})}\BibitemShut {NoStop}%
\bibitem [{\citenamefont {Schraft}\ \emph {et~al.}(2018)\citenamefont
  {Schraft}, \citenamefont {Junginger}, \citenamefont {Feldmaier},
  \citenamefont {Bardakcioglu}, \citenamefont {Main}, \citenamefont {Wunner},\
  and\ \citenamefont {Hernandez}}]{hern18c}%
  \BibitemOpen
  \bibfield  {author} {\bibinfo {author} {\bibfnamefont {P.}~\bibnamefont
  {Schraft}}, \bibinfo {author} {\bibfnamefont {A.}~\bibnamefont {Junginger}},
  \bibinfo {author} {\bibfnamefont {M.}~\bibnamefont {Feldmaier}}, \bibinfo
  {author} {\bibfnamefont {R.}~\bibnamefont {Bardakcioglu}}, \bibinfo {author}
  {\bibfnamefont {J.}~\bibnamefont {Main}}, \bibinfo {author} {\bibfnamefont
  {G.}~\bibnamefont {Wunner}}, \ and\ \bibinfo {author} {\bibfnamefont
  {R.}~\bibnamefont {Hernandez}},\ }\href {\doibase 10.1103/PhysRevE.97.042309}
  {\bibfield  {journal} {\bibinfo  {journal} {Phys. Rev. E}\ }\textbf {\bibinfo
  {volume} {97}},\ \bibinfo {pages} {042309} (\bibinfo {year}
  {2018})}\BibitemShut {NoStop}%
\bibitem [{\citenamefont {Craven}, \citenamefont {Bartsch},\ and\ \citenamefont
  {Hernandez}(2014)}]{hern14f}%
  \BibitemOpen
  \bibfield  {author} {\bibinfo {author} {\bibfnamefont {G.~T.}\ \bibnamefont
  {Craven}}, \bibinfo {author} {\bibfnamefont {T.}~\bibnamefont {Bartsch}}, \
  and\ \bibinfo {author} {\bibfnamefont {R.}~\bibnamefont {Hernandez}},\ }\href
  {\doibase 10.1063/1.4891471} {\bibfield  {journal} {\bibinfo  {journal} {J.
  Chem. Phys.}\ }\textbf {\bibinfo {volume} {141}},\ \bibinfo {pages} {041106}
  (\bibinfo {year} {2014})}\BibitemShut {NoStop}%
\bibitem [{\citenamefont {Revuelta}\ \emph {et~al.}(2017)\citenamefont
  {Revuelta}, \citenamefont {Craven}, \citenamefont {Bartsch}, \citenamefont
  {Borondo}, \citenamefont {Benito},\ and\ \citenamefont
  {Hernandez}}]{hern17f}%
  \BibitemOpen
  \bibfield  {author} {\bibinfo {author} {\bibfnamefont {F.}~\bibnamefont
  {Revuelta}}, \bibinfo {author} {\bibfnamefont {G.~T.}\ \bibnamefont
  {Craven}}, \bibinfo {author} {\bibfnamefont {T.}~\bibnamefont {Bartsch}},
  \bibinfo {author} {\bibfnamefont {F.}~\bibnamefont {Borondo}}, \bibinfo
  {author} {\bibfnamefont {R.~M.}\ \bibnamefont {Benito}}, \ and\ \bibinfo
  {author} {\bibfnamefont {R.}~\bibnamefont {Hernandez}},\ }\href {\doibase
  10.1063/1.4997571} {\bibfield  {journal} {\bibinfo  {journal} {J. Chem.
  Phys.}\ }\textbf {\bibinfo {volume} {147}},\ \bibinfo {pages} {074104}
  (\bibinfo {year} {2017})}\BibitemShut {NoStop}%
\bibitem [{\citenamefont {Tsch{\"o}pe}\ \emph {et~al.}(2020)\citenamefont
  {Tsch{\"o}pe}, \citenamefont {Feldmaier}, \citenamefont {Main},\ and\
  \citenamefont {Hernandez}}]{hern20d}%
  \BibitemOpen
  \bibfield  {author} {\bibinfo {author} {\bibfnamefont {M.}~\bibnamefont
  {Tsch{\"o}pe}}, \bibinfo {author} {\bibfnamefont {M.}~\bibnamefont
  {Feldmaier}}, \bibinfo {author} {\bibfnamefont {J.}~\bibnamefont {Main}}, \
  and\ \bibinfo {author} {\bibfnamefont {R.}~\bibnamefont {Hernandez}},\ }\href
  {\doibase 10.1103/PhysRevE.101.022219} {\bibfield  {journal} {\bibinfo
  {journal} {Phys. Rev. E}\ }\textbf {\bibinfo {volume} {101}},\ \bibinfo
  {pages} {022219} (\bibinfo {year} {2020})}\BibitemShut {NoStop}%
\bibitem [{\citenamefont {Goldstein}(1980)}]{goldstein80}%
  \BibitemOpen
  \bibfield  {author} {\bibinfo {author} {\bibfnamefont {H.}~\bibnamefont
  {Goldstein}},\ }\href {https://books.google.de/books?id=I1JKjwEACAAJ} {\emph
  {\bibinfo {title} {Classical Mechanics}}},\ Addison-Wesley Series in Physics\
  (\bibinfo  {publisher} {Addison--Wesley},\ \bibinfo {address} {London},\
  \bibinfo {year} {1980})\BibitemShut {NoStop}%
\end{thebibliography}%

\end{document}


\title{Supplementary material for
``Influence of external driving on decays in the geometry of the LiCN isomerization''}
\author{Matthias Feldmaier}
\affiliation{%
Institut f\"ur Theoretische Physik 1,
Universit\"at Stuttgart,
70550 Stuttgart,
Germany}
\author{Johannes Reiff}
\affiliation{%
Institut f\"ur Theoretische Physik 1,
Universit\"at Stuttgart,
70550 Stuttgart,
Germany}
\author{Rosa M. Benito}
\affiliation{Grupo de Sistemas Complejos,
  Escuela T\'ecnica Superior de Ingenier\'ia Agron\'omica,
        Alimentaria y de Biosistemas,
  Universidad Polit\'ecnica de Madrid,
  28040 Madrid, Spain}
\author{Florentino Borondo}
\affiliation{Instituto de Ciencias Matem\'aticas (ICMAT),
  Cantoblanco, 28049 Madrid, Spain}
\affiliation{Departamento de Qu\'{\i}mica,
Universidad Aut\'onoma de Madrid,
Cantoblanco, 28049 Madrid, Spain}
\author{J\"org Main}
\affiliation{%
Institut f\"ur Theoretische Physik 1,
Universit\"at Stuttgart,
70550 Stuttgart,
Germany}

\author{Rigoberto Hernandez}
\email[Correspondence to: ]{r.hernandez@jhu.edu}
\affiliation{%
Department of Chemistry,
Johns Hopkins University,
Baltimore, Maryland 21218, USA}
\affiliation{%
    Departments of Chemical \& Biomolecular Engineering,
    and Materials Science and Engineering,
    Johns Hopkins University,
    Baltimore, Maryland 21218, USA
}

\date{\today}

\maketitle

\preto\section\acresetall
\acrodef{DS}{dividing surface}
\acrodef{NHIM}{normally hyperbolic invariant manifold}
\acrodef{PSOS}{Poincar\'e surface of section}
\acrodef{NN}{neural network}
\acrodef{TST}{transition state theory}
\acrodef{TS}{transition state}
\acrodef{LD}{Lagrangian descriptor}
\acrodef{TD}{time descriptor}
\acrodef{BCM}{binary contraction method}
\acrodef{EOM}{equations of motion}
\acrodef{LMA}{local manifold analysis}
\acrodef{SCF}{self-consistent field}


\section{Contents}
In this supplementary material, we provide a detailed
description of the methods used in the primary text.
In Sec.~\ref{sec:decay_rates}, we summarize the
details for the implementation of three approaches
in determining the decay rates: Floquet analysis, ensemble method and
the local manifold analysis.
A technical derivation of the last of these methods is
included separately in Sec.~\ref{sec:lma}.

\section{Decay rates of reactant population close to the TS}
\label{sec:decay_rates}
\begin{figure}
\includegraphics[width=\columnwidth]{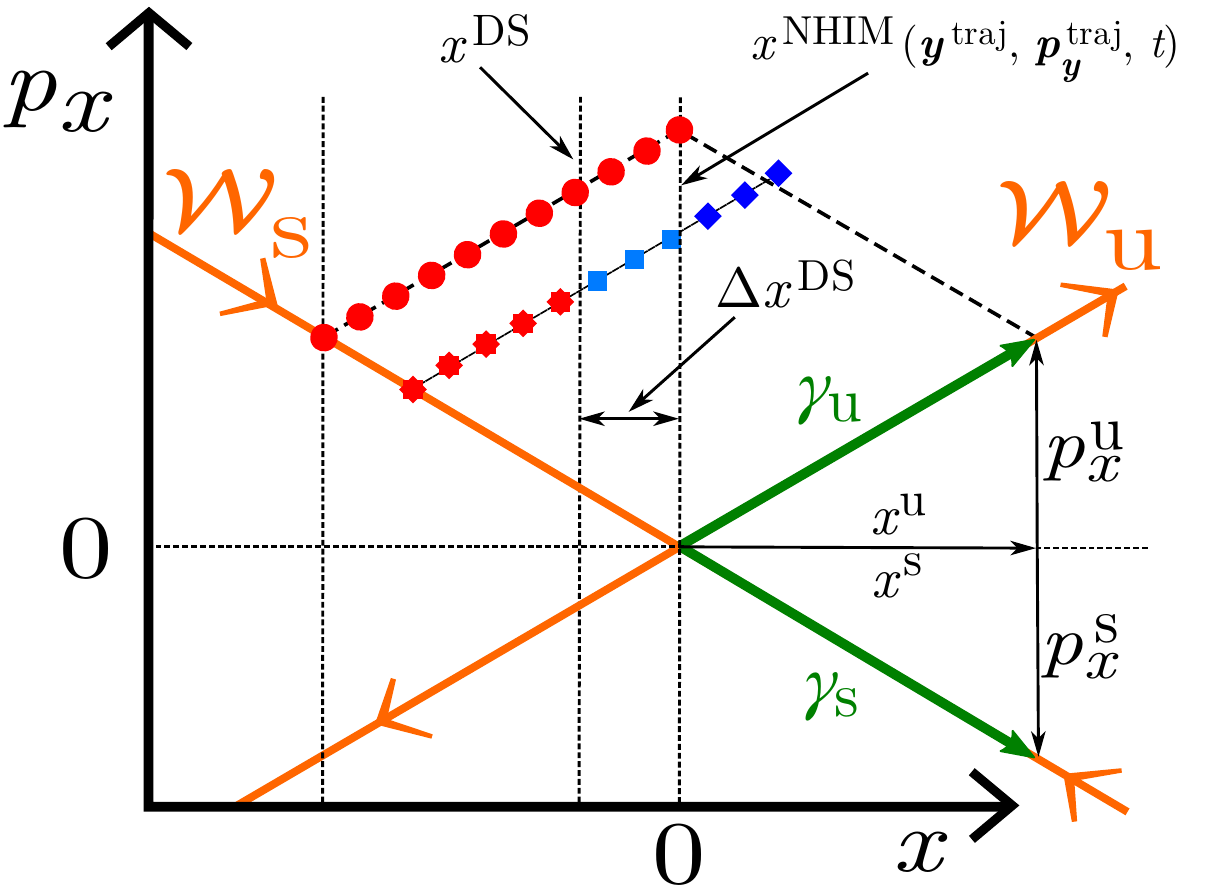}
    \caption{%
        Schematic $(x, p_x)$-section of the phase space in a close neighborhood
        $\mleft(\vec{y}^{\no{traj}}, \vec{p_y}^{\no{traj}}, t\mright)$ of
        a specific trajectory on the \ac{NHIM}.
Two contributions of a decaying reactant population to
        a decay rate are highlighted.
One is based on the ensemble moving with respect to
        the trajectory on the \ac{NHIM} (dark blue diamonds),
and the other is based
        on a \emph{relative movement} of the \ac{DS} for an ensemble which turns
        out of this $(x, p_x)$ plane (light blue squares).
    }
    \label{fig:lma_together}
\end{figure}

The phase space of a system with $n$ degrees of freedom and a rank-1
saddle can be described, at least in a local neighborhood of the NHIM,
by one reaction coordinate $x$ and the
corresponding momentum $p_x$, $n-1$ orthogonal modes $\vb{y}$, and the
corresponding momenta $\vb{p}_y$.
In LiNC (see Fig.~1
), for example,
these can be represented using
the angle $\vartheta$ as the reaction coordinate $x$,
and $R$ as the orthogonal mode $y$
as long as one stays near the reaction path.
For such a system,
trajectories are classified as ``reactive'' if they are able to overcome the rank-1 barrier
between two separate regions of the potential energy surface
referred to as reactants $\mathcal{R}$ and products $\mathcal{P}$.
Trajectories, which cannot overcome the barrier between these clearly
distinct regions are considered ``non-reactive''.
For a fixed position $\vb{y}$ and $\vb{p}_y$ (and at a given time
$t$), a corresponding cross section $(x, p_x)$ of the phase space in the barrier region
is illustrated in Fig.~\ref{fig:lma_together}.
Usually, such a cross-section is divided into four disjunct regions
for which two correspond
to reactive trajectories and the other two to non-reactive trajectories.\cite{hern17e,hern19a}
These four regions are separated by the stable and unstable manifolds
marked $\mathcal{W}\sno{s}$ and $\mathcal{W}\sno{u}$ in
Fig.~\ref{fig:lma_together}.
Their intersection yields a point $(x, p_x)^\no{NHIM}(\vb{y}, \vb{p}_y, t)$ of the \ac{NHIM}.
The \ac{NHIM} is an unstable co-dimension two manifold containing all the trajectories
that (at least mathematically) stay in the barrier region when being propagated
both forward and backward in time.\cite{hern19a}
For given $(\vb{y}, \vb{p}_y, t)$, the corresponding position $(x,p_x)^{\mathrm{NHIM}}$
of the \ac{NHIM} can be computed using the \ac{BCM}.\cite{hern18g}
At this position, a vertical \ac{DS} can be attached,
as illustrated by the vertical dashed line
in Fig.~\ref{fig:lma_together} which is recrossing-free at least in a close neighborhood
of the \ac{NHIM}.\cite{hern19a,hern19e,hern18c}

In this paper, we investigate the dynamics and decay rates of the
structures in the close vicinity of the \ac{TS}
in a driven LiCN isomerization reaction.
We have developed various numerical methods\cite{hern19e}
for the computation of the decay rates of reactant populations of
multidimensional driven systems with a rank-1 saddle---\emph{viz.}\ the
ensemble method, the \acf{LMA}, and the Floquet method.
For the convenience of the reader we briefly recapitulate the
basic ideas of these methods, which are then applied in
the results section to the LiCN isomerization reaction.

\subsection{Floquet method for average decay rates}
\label{sec:Floquet_method}

Reaction rates of the \ac{TS} in a one-dimensional system had
earlier been obtained directly from the moving \ac{TS} trajectory through
its stability exponents.\cite{hern14f,hern17f}
The latter correspond to Floquet exponents in a periodically driven
system,\cite{hern14f}
and Lyaponov exponents in nonperiodic driving.\cite{hern17f}
In these early examples, the decay rates associated with the
\ac{TS} trajectory were identifiable as reaction rates because
the local \ac{TS} geometry was actually global in the sense of
identifying long-time reactivity and global non-recrossing.
However, for more general energy landscapes, the determination
of the \ac{TS} trajectory in 1-dimension or its generalizations
may be local in the sense of being obtained within a limited
domain associated with a barrier (or barriers.)
This is the sense in which we are obtaining local decay rates
and not necessarily global reaction rates in this work.
Thus, the stability of the \ac{TS} trajectory obtained
within a given local region is associated with
the so-called Floquet rate which 
uses the Floquet exponents to obtain
the reaction rate
when the local geometry is global.

In multidimensional cases, one has to take care of the fact that
the \ac{TS} trajectory becomes a time-dependent
codimension-2 manifold---viz, the \ac{NHIM}---associated
with infinitely many trajectories which we refer to as bound
trajectories.\cite{hern19e,hern20d}
For periodically driven arbitrary-dimensional systems,
the decay rates (or Floquet rates), are simply obtained by the difference
\begin{equation}
    \label{eq:floquet_rate}
    k\sno{F} = \mu\sno{l} - \mu\sno{s}
    \MathComma
\end{equation}
between a Floquet exponent $\mu\sno{l}$ associated with the unstable direction
of trajectories moving away from the \ac{NHIM},
and a Floquet exponent $\mu\sno{s}$ associated with
the stable motion of the trajectories towards the \ac{NHIM}.
In 1-dimension, there are only two such exponents, and hence
there is only one choice for $\mu\sno{l}$ and $\mu\sno{s}$.\cite{hern14f}
In general, however, there are many such exponents and infinitely
many bound trajectories.
Nevertheless, the structure of Eq.~\eqref{eq:floquet_rate}
holds as specified below.

The Floquet exponents are a measure of how exponentially fast initially
neighboring trajectories separate from each other.\cite{goldstein80}
Consequently, they report the stability of these trajectories.
The dynamics in the direction of the $n - 1$ orthogonal modes
of a rank-1 saddle is \emph{stable}.
Initially neighboring trajectories do not separate exponentially fast
as they are propagated
with respect to these stable degrees of freedom%
---as long as the dynamics inside the \ac{NHIM} is not chaotic.
Consequently, the corresponding Floquet exponents vanish
and the related eigenvalues
of the monodromy matrix have an absolute value of one.
In the unstable direction of a rank-1 barrier, however,
initially neighboring trajectories will depart exponentially
fast at least in a local neighborhood of the \ac{NHIM} and,
hence, the associated Floquet exponents do not vanish.
For a periodic trajectory with period $T$ on the \ac{NHIM},
these non-vanishing Floquet exponents are obtained
from
\begin{equation}
    \label{eq:floquet_exponents}
    \mu\sno{l, s} = \frac{1}{T} \ln|m\sno{l, s}|\MathPeriod
\end{equation}
Here, $m\sno{l}$ is the largest
and $m\sno{s}$ is the smallest eigenvalue of the corresponding
monodromy matrix $\vec{M}(T)$,
and this specifies precisely the meaning of the two chosen
Floquet exponents in Eq.~\eqref{eq:floquet_rate}.
The monodromy matrix $\vec{M}(T) = \vec{\sigma}(t_0 + T, t_0)$ is a
special case of the fundamental matrix $\vec{\sigma}(t, t_0)$, which is
obtained by solving the set of differential equations
\begin{equation}
    \label{eq:lma_eom_fund_mat}
    \dv{t} \vec{\sigma}(t, t_0)
    = \eval{\pdv{\dot{\gamma}_i}{\gamma_j}}_{\vec{\gamma}(t)} \vec{\sigma}(t, t_0)
    ~\,\no{with}~\,
    \vec{\sigma}(t_0, t_0)
    = \mathbbm{1}
    \MathPeriod
\end{equation}
Here, $\dot{\vec{\gamma}}$ are the $(2n)$-dimensional equations of
motion in phase space and $\vec{\gamma}(t)$ is a trajectory on the
\ac{NHIM}.
Since the Floquet exponents
are obtained for a full period of the external driving
according to Eqs.~\eqref{eq:floquet_exponents},
the Floquet rate~\eqref{eq:floquet_rate} is an integrated quantity corresponding
to a mean decay rate averaged over a full periodic trajectory.

According to Ref.~\onlinecite{hern19e},
only a small set (of measure zero) of trajectories on the
\ac{NHIM} of a periodically driven system are periodic, but 
nearly all of them do show quasi-periodic behavior as long as the
dynamics in the orthogonal modes is not chaotic.
For these quasi-periodic trajectories
inside the \ac{NHIM},
Eq.~\eqref{eq:floquet_exponents} needs to be modified as
\begin{equation}
    \label{eq:floquet_exponents_non_periodic}
    \mu\sno{l, s} = \lim_{t \to \infty} \frac{1}{t} \ln|m\sno{l, s}(t)|
\end{equation}
to obtain the Floquet exponents associated with a specific torus on the \ac{NHIM}.
Since the calculation of the infinite time limit
in Eq.~\eqref{eq:floquet_exponents_non_periodic}
is numerically impossible,
the integration time $t$ must, in practice, be taken to
be sufficiently large to cover
the relevant or characteristic dynamics of the observed
trajectory---e.\,g., long enough to traverse several windings on a stable torus.
In doing so, the Floquet rate of such a quasi-periodic
trajectory corresponds to the mean decay rate of
the reactant population close to the respective torus.

\subsection{Ensemble method for instantaneous decay rates}
\label{sec:ensemble_method}

The ensemble method is an
intuitive numerical method for obtaining the instantaneous decay rate
of a reactant population close to the \ac{TS}.
Here, a homogeneous and linear ensemble
of $N\sno{react}$ reactive trajectories is initialized on the reactant side
of the full phase space.
Specifically, they are placed on an $(x, p_x)$-cross sectional surface
at a small distance $\Delta x$
from a given position $(\vec{y}^\no{traj}(t), \vec{p_y}^\no{traj}(t))$
of an arbitrarily chosen trajectory
on the \ac{NHIM} (see red bullets in Fig.~\ref{fig:lma_together}).
After propagating this ensemble for a time $\Delta t$,
a subdomain
will have pierced the \ac{DS} and entered the product side
(dark blue diamonds and light blue squares in Fig.~\ref{fig:lma_together})
while the remainder will still be
located on the reactant side (red stars in Fig.~\ref{fig:lma_together}).
As the \ac{DS} is non-recrossing, the
resulting decrease in the reactant population of a close neighborhood of the
\ac{NHIM} is associated with a rate
\begin{equation}
    \label{eq:inst_ensemble_rate}
    k\sno{e}(\vb{y}, \vb{p}_y, t)
    = -\dv{t} \ln(N\sno{react}(\vb{y}, \vb{p}_y, t))
    \MathPeriod
\end{equation}
It is referred to as the \emph{instantaneous ensemble rate}
to emphasize that it is obtained by propagating an ensemble of reactive trajectories
according to the equations of motion. In doing so,
the \ac{DS} is computed individually for each trajectory of
the ensemble and each time step.

The ensemble can be propagated not just for a small time
step $\Delta t$ but for longer times when computing the time-dependent
ensemble rate according to Eq.~\eqref{eq:inst_ensemble_rate}.
With increasing time, however,
the initial ensemble becomes more and more
distorted (see Sec.~\ref{sec:lma} for further information).
Since the \ac{DS} is computed individually for each
reactive trajectory, such distortion effects are automatically taken into account
using the ensemble method.
As the reactant population also decreases exponentially when propagating the ensemble,
a new ensemble can be initialized close to the
corresponding point of the respective trajectory on the \ac{NHIM}
after an appropriately chosen propagation time.
Such technical details are discussed in Ref.~\onlinecite{hern19e}.
Although the implementation of the ensemble method is straightforward,
it can be numerically expensive
because the ensemble consists of many trajectories and
the \ac{DS} is obtained individually for each reactive trajectory
using the \ac{BCM}.\cite{hern18g}

\subsection{Local manifold analysis}
\label{sec:lma}

We can avoid most of the expensive particle propagation of the
ensemble method by leveraging
the geometry of the stable and unstable manifolds
to effectively describe
the linearized dynamics near the \ac{NHIM}.
The resulting \ac{LMA} method\cite{hern19e}
can thus be seen as an extension to the ensemble method
with the difference that the time when
trajectories have reached the \ac{DS}
can now be determined analytically through the linearization
thereby avoiding a costly numerical integration.
As a result,
the computational effort required to calculate instantaneous decay rates
is significantly reduced
while numerical precision is simultaneously enhanced.

Similar to the ensemble method, in the \ac{LMA}, we consider
the region of phase space close to a trajectory
$\gammaTS = (\xTS, \yTS, \pxTS, \pyTS)^\transpose$
on the \ac{NHIM},
as shown in Fig.~\ref{fig:lma_together}.
For simplicity, we choose a
moving coordinate frame in which the origin is at
$\xTS = \pxTS = 0$ for all times $t$.

The decay rate is determined by two components:

(i) The first contribution arises from the movement
of the ensemble akin to Sec.~\ref{sec:ensemble_method}
relative to \gammaTS.
The resulting flux through the associated \ac{DS} at $\xDS = 0$
(dark blue diamonds in Fig.~\ref{fig:lma_together})
is then obtained via the slopes of the stable and unstable manifolds
defined by the variables $\xS = \xU$, \pxS, and \pxU.
The details and mathematical underpinnings of this procedure,
based on Ref.~\onlinecite{hern19e},
may be found in Sec.~\ref{sec:lma_mathematical}.

(ii) The second contribution accounts for the fact that
in systems with more than one degree of freedom,
the ensemble can turn out of the $(x, p_x)$ plane associated with \gammaTS.
This can happen
if the system's orthogonal modes are coupled
to the reaction coordinate momentum $p_x$
and leads to an apparent movement of the \ac{DS} relative to \gammaTS\
(represented by $\Delta \xDS$ and bright blue diamonds
in Fig.~\ref{fig:lma_together}).
As a result, the instantaneous flux through the \ac{DS} is modified.
To quantify this effect,
we first propagate the top particle of the ensemble
initially located on the \ac{DS}
numerically by a small time step $\delta t$.
The related shift of the \ac{DS}, $\xDS(t + \var{t})$,
can then be determined by
projecting the propagated particle back onto the \ac{NHIM} using the
\ac{BCM}.

Combining the two terms,
the instantaneous decay rate can be written as
\begin{equation}
    \label{eq:k_lma}
    \kM(t; \gammaTS)
    = \mat{J}_{x, p_x}(t) \frac{\pxU(t) - \pxS(t)}{\xU(t)}
        - \frac{\xDS(t + \var{t})}{\xU(t) \var{t}}
    \MathComma
\end{equation}
where $\mat{J}(t)$ is the Jacobian of the system's equations of motion
evaluated for the trajectory \gammaTS\ at time $t$.
A more detailed derivation of Eq.~\eqref{eq:k_lma}
is provided in Sec.~\ref{sec:lma_mathematical}

\section{Derivation of the local manifold analysis}
\label{sec:lma_mathematical}

We consider a trajectory
$\gammaTS(t) = (\xTS, \yTS, \pxTS, \pyTS)^\transpose$
starting at some arbitrary time $t_0$ on the \ac{NHIM}.
All of the statements in this section
depend implicitly on \gammaTS\ and $t_0$,
which we neglect in our notation for simplicity.
Without loss of generality,
we choose coordinates such that $\xTS(t) = \pxTS(t) = 0$ for all times $t$.
Figure~\ref{fig:lma_together} sketches
an $(x, p_x)$-section of phase space in close proximity to $\gammaTS(t_0)$.
We can assume that the manifold fibers in this section are straight lines
since we only look at the dynamics very close to the \ac{NHIM}.
Therefore, the stable and unstable manifolds can be described
using only two vectors
$\gammaS = (\xS, \pxS)^\transpose$ and $\gammaU = (\xU, \pxU)^\transpose$.
These vectors will be squeezed and stretched as a function of time
if subjected to the equations of motion.
Without loss of generality,
we initially choose $0 < \xS(t_0) = \xU(t_0)$
such that we are in the linear regime.

To obtain a decay rate for $\gammaTS(t_0)$, we now consider
a linear, equidistant ensemble parameterized by
\begin{equation}
    \label{eq:lma/ensemble}
    \gammaE(a, t) = -\gammaS(t) + a \gammaU(t)
\end{equation}
where $a \in [0, 1]$.
The ensemble is constructed parallel to the
unstable manifold---see Fig.~\ref{fig:lma_together}.
Initially, the ensemble pierces the \ac{DS} at $\aDS(t_0) = 1$
(circles in Fig.~\ref{fig:lma_together}).
As time goes by, however, the ensemble will be stretched
and $\aDS(t)$ will therefore decay exponentially
(diamonds in Fig.~\ref{fig:lma_together}).
More precisely, \aDS\ is proportional to the number of reactants
and therefore leads to a decay rate
\begin{equation}
    \label{eq:lma/decay}
    \kM(t_0)
    = -\eval{\dv{t} \ln(\aDS(t))}_{t_0}
    = -\aDSdot(t_0)
\end{equation}
at time $t_0$ in analogy to Eq.~\eqref{eq:inst_ensemble_rate}.
In this picture, the total decay rate consists of two contributions
\begin{equation}
    \label{eq:lma/parts}
    \kM(t_0) = \kEns(t_0) + \kDS(t_0)
    \MathPeriod
\end{equation}

For the first part, \kEns, we assume that
the ensemble stays in the $(x, p_x)$-section associated with $\gammaTS(t)$.
As a result, the point where the ensemble pierces the \ac{DS}
is fixed at $\xDS(t) = 0$ for all times $t$.
This is an effectively one-dimensional model.
We start by looking at the linearized dynamics near the \ac{NHIM}
\begin{equation}
    \dv{t} \vec{\gamma}(t) = \mat{J}(t) \vec{\gamma}(t)
    \MathComma
\end{equation}
where $\mat{J}(t)$ is the Jacobian of the system's equations of motion
evaluated on the trajectory \gammaTS.
The fundamental matrix $\mat{\sigma}(t)$
obtained by integrating $\dot{\mat{\sigma}}(t) = \mat{J}(t) \mat{\sigma}(t)$
with $\mat{\sigma}(t_0) = \mathbbm{1}$
[\cf\ Eq.~\eqref{eq:lma_eom_fund_mat}]
can then be used to propagate the ensemble
from time $t_0$ to a later time $t$ via
\begin{equation}
    \gammaE(a, t) = \mat{\sigma}(t) \gammaE(a, t_0)
    \MathPeriod
\end{equation}
We are interested in the point $\aDS$ where $\gammaE(a, t)$ pierces the \ac{DS}
at $\xDS(t) = 0$,
\ie,
\begin{equation}
    \mat{\sigma}(t) \gammaE(\aDS(t), t_0) \vdot \vu{e}_x \overset{!}{=} 0
    \MathPeriod
\end{equation}
Inserting Eq.~\eqref{eq:lma/ensemble} yields
\begin{equation}
    \aDS(t) = \frac
        {\mat{\sigma}_{x, x}(t) \xU(t_0) + \mat{\sigma}_{x, p_x}(t) \pxS(t_0)}
        {\mat{\sigma}_{x, x}(t) \xU(t_0) + \mat{\sigma}_{x, p_x}(t) \pxU(t_0)}
    \MathComma
\end{equation}
where we have used $\xS(t_0) = \xU(t_0)$.
This intermediate result can be substituted into Eq.~\eqref{eq:lma/decay}.
Since we are only interested in the instantaneous rate at $t = t_0$,
we can simplify the expression
using $\mat{\sigma}(t_0) = \mathbbm{1}$
as well as $\dot{\mat{\sigma}}_{x, p_x}(t_0) = \mat{J}_{x, p_x}(t_0)$
and arrive at
\begin{equation}
    \label{eq:lma/ens}
    \kEns(t_0)
    = \mat{J}_{x, p_x} \frac{\pxU(t_0) - \pxS(t_0)}{\xU(t_0)}
    \MathPeriod
\end{equation}
A geometric interpretation of \kEns\ can be found in Ref.~\onlinecite{hern19e}.

The second contribution, \kDS, in Eq.~\eqref{eq:lma/parts} stems from
the fact that in systems with more than one degree of freedom
the ensemble may leave the $(x, p_x)$-section associated with $\gammaTS(t)$.
An ensemble moving out-of-plane can mostly be treated as described above
by projecting it back onto the $(x, p_x)$-section.
Since the position of the \ac{DS} $\xDS(\vec{y}, \vec{p}_y)$
is dependent on the orthogonal modes, however,
this may lead to the ensemble intersecting with the \ac{DS} at $\xDS \ne 0$.

To quantify the effect on \kM, consider a small time step $\var{t}$.
In the linear regime, the change $\var{a}$ caused
by the ensemble drifting out-of-plane
can be written as
\begin{equation}
    \var{\aDS(t_0)}
    = \frac{\xDS(t_0 + \var{t}) - \xDS(t_0)}{\xU(t_0)}
    \MathComma
\end{equation}
where $\xU(t_0)$ accounts for normalization.
Using Eq.~\eqref{eq:lma/decay} and the fact that $\xDS(t_0) = 0$, we obtain
\begin{equation}
    \label{eq:lma/ds}
    \kDS(t_0)
    = -\fdv{\aDS(t_0)}{t}
    = -\frac{\xDS(t_0 + \var{t})}{\xU(t_0) \var{t}}
    \MathPeriod
\end{equation}
The quantities $\var{t}$ and \xU\ can be freely chosen within certain limits,
while $\xDS(t_0 + \var{t})$ can be determined numerically by
propagating the particle $\gammaE(1, t_0)$ initially located on the \ac{DS}
for $\var{t}$ units of time
and projecting it back onto the \ac{NHIM}
using the \ac{BCM}.\cite{hern18g}
By combining Eqs.~\eqref{eq:lma/ens} and~\eqref{eq:lma/ds}
according to Eq.~\eqref{eq:lma/parts},
we finally arrive at the instantaneous decay rate $\kM(t; \gammaTS)$
for a trajectory \gammaTS\ on the \ac{NHIM} given in Eq.~\eqref{eq:k_lma}
of Sec.~\ref{sec:lma}.

\section*{References}
\bibliography{SM-q21}